\documentclass[journal, twoside]{IEEEtran}
\usepackage{cite}

\ifCLASSINFOpdf
  \usepackage[pdftex]{graphicx}
\else
\fi
\usepackage{amsmath}
\usepackage{algorithmic}
\usepackage{algorithm}

\usepackage{array}
\usepackage{url}

\usepackage{booktabs} 
\usepackage{multirow}
\usepackage{xcolor} 
\usepackage[super]{nth}
\usepackage{tikz}
\newcommand*\circled[1]{\tikz[baseline=(char.base)]{
            \node[shape=circle,draw,inner sep=0.2pt] (char) {#1};}}
\newcommand*\circledB[1]{\tikz[baseline=(char.base)]{
            \node[shape=circle,fill,inner sep=0.2pt] (char) {\textcolor{white}{#1}};}}
\usepackage{soul}
\usepackage{enumitem}
\newcommand{\add}[1]{\textcolor{black}{#1}}
\newcommand{\addx}[1]{\textcolor{black}{#1}}

\usepackage[none]{hyphenat}

\begin{document}
%
\title{\huge ROMANet: Fine-Grained Reuse-Driven Off-Chip Memory Access Management and Data Organization \\ for Deep Neural Network Accelerators}
%
%
%
\author{Rachmad~Vidya~Wicaksana~Putra, Muhammad~Abdullah~Hanif,~\IEEEmembership{Student~Member,~IEEE,} Muhammad~Shafique,~\IEEEmembership{Senior~Member,~IEEE}
\thanks{R. V. W. Putra, M. A. Hanif, and M. Shafique are with the Institute of Computer Engineering, Technische Universit\"at Wien, Austria. E-mail: (rachmad.putra, muhammad.hanif, muhammad.shafique)@tuwien.ac.at}
\vspace{-0.8cm}
}

\markboth{Journal of \LaTeX\ Class Files,~Vol.~xx, No.~x, 2020}%
{Putra \MakeLowercase{\textit{et al.}}: ROMANet: Reuse-Driven Off-Chip Memory Access Management for DNN Accelerators}
%



\maketitle

\vspace{-0.4cm} 
\begin{abstract}
Enabling high energy efficiency is crucial for embedded implementations of deep learning.
Several studies have shown that the DRAM-based off-chip memory accesses are one of the most energy-consuming operations in deep neural network (DNN) accelerators, and thereby limit the designs from achieving efficiency gains at the full potential. 
DRAM access energy varies 
depending upon the number of accesses required as well as the energy consumed per-access. 
Therefore, searching for a solution towards the minimum DRAM access energy is an important optimization problem. 
Towards this, we propose the ROMANet methodology that aims at reducing the number of memory accesses, by searching for the appropriate data partitioning and scheduling for each layer of a network using a design space exploration, based on the knowledge of the available on-chip memory and the data reuse factors. 
\add{Moreover, ROMANet also targets decreasing the number of DRAM row buffer conflicts and misses, by exploiting the DRAM multi-bank burst feature to improve the energy-per-access. 
Besides providing the energy benefits, our proposed DRAM data mapping also results in an increased effective DRAM throughput, which is useful for latency-constraint scenarios.}
Our experimental results show that the ROMANet saves DRAM access energy by 12\% for the AlexNet, by 36\% for the VGG-16, and by 46\% for the MobileNet, \add{while also improving the DRAM throughput by 10\%}, as compared to the state-of-the-art. 
\end{abstract}

\vspace{-0.1cm}
\vspace{-0.3cm}

%
\IEEEpeerreviewmaketitle

\section{Introduction}
%
%
%
%
\label{Sec:Intro}
\IEEEPARstart{C}{}onvolutional neural networks (CNNs), a particular type of deep neural networks (DNNs), have emerged as a promising solution for a wide range of machine learning applications, e.g., image classification, object detection, autonomous vehicles, and smart health-care \cite{Ref:LeCun_DeepLearning_Nature15}. 
To expedite the inference process, several CNN hardware accelerators have been proposed over the past few years\cite{Ref:Chen_DianNao_ASPLOS14,Ref:Zhang_CNNfpga_FPGA15,Ref:Han_EIE_ISCA16, Ref:Chen_Eyeriss_JSSC16, Ref:Zhang_CambriconX_MICRO15, Ref:Albericio_Cnvlutin_ISCA16, Ref:Luo_DaDianNao_TC17, Ref:Jouppi_TPU_ISCA17, Ref:Parashar_SCNN_ISCA17, Ref:Lu_FlexFlow_HPCA17, Ref:Kwon_MAERI_ASPLOS18, Ref:Hanif_MPNA_arXiv18, Ref:Sharma_BitFusion_ISCA18}. 
These accelerators offer a higher performance efficiency as compared to the general-purpose CPU-based solution. 
However, most of them only present \textit{isolated} accelerator design, and do not thoroughly study the impact of off-chip memory accesses, 
especially for scenarios where \textit{the full CNN processing cannot be mapped at the same time on an accelerator fabric}. 
The reason is that the CNNs are large in size, and typical compute engine and on-chip memory are small and may not even be sufficient to process one complete layer of a network at a time. 
Moreover, each data is usually involved in multiple computations, i.e., multiply-and-accumulate (MAC) operations, during processing. 
Therefore, multiple 
redundant accesses for the same data to off-chip memory are inevitable. 
\begin{figure}[t]
\centering
\includegraphics[width=\linewidth]{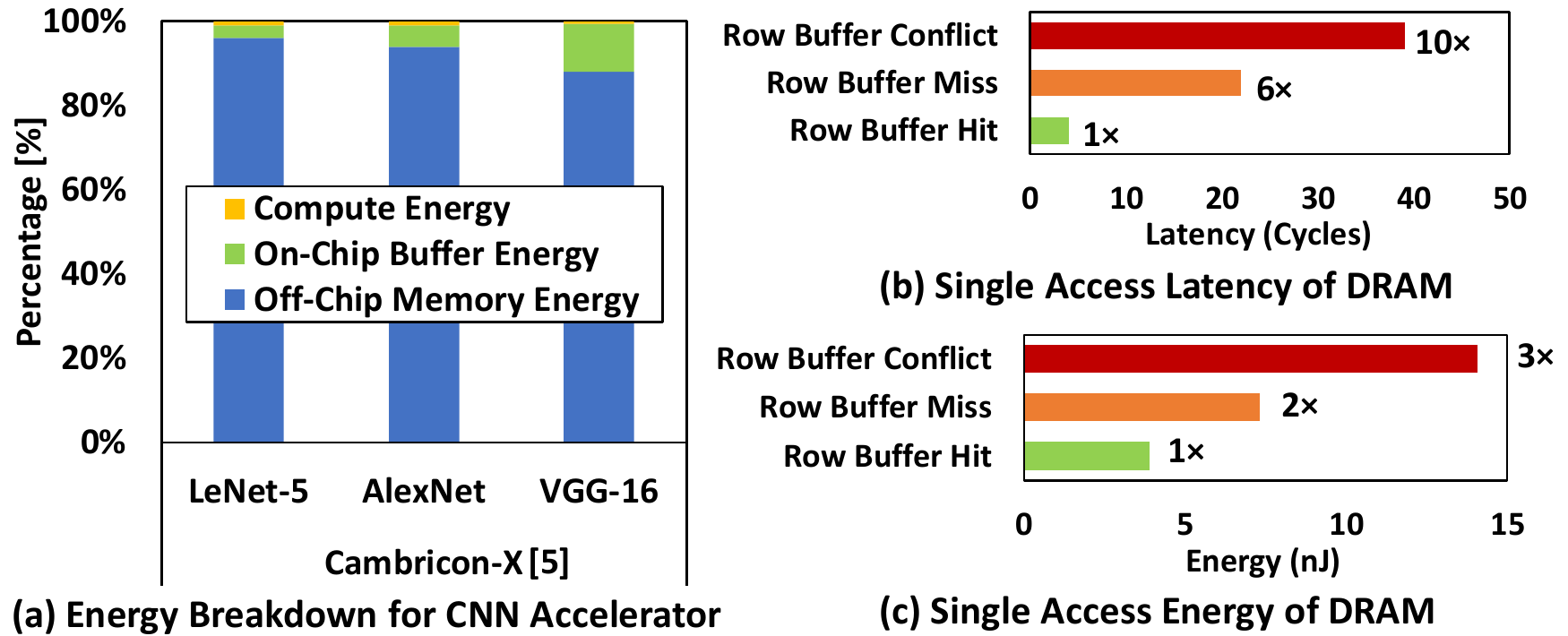}
\vspace{-0.7cm}
\caption{(a) Breakdown of the total energy consumption of CNN accelerator, i.e., Cambricon-X \cite{Ref:Zhang_CambriconX_MICRO15}. DRAM access energy consumes $>80\%$ of the total energy. Profile of (b) DRAM access latency and (c) DRAM access energy for a row buffer conflict, miss, and hit. Data are obtained from our experiments for DDR3-1600 2Gb x8 using cycle-accurate DRAM simulators \cite{Ref:Kim_Ramulator_LCA15} \cite{Ref:Ghose_VAMPIRE_POMACS18}.}
\label{Fig:BreakdownLatencyEnergyDRAM}
\vspace{-0.6cm}
\end{figure}
The redundant accesses to DRAM-based off-chip memory hinder the accelerators from achieving further efficiency gains as DRAM energy is significantly higher compared to other operations \cite{Ref:Sze_DNNsurvey_IEEE17, Ref:Horowitz_ComputeEnergy_ISSCC14, Ref:Hedge_UCNN_ISCA18}, which is also shown in Fig. \ref{Fig:BreakdownLatencyEnergyDRAM}(a).
Therefore, savings in DRAM energy can result in proportional system-level energy savings. 

DRAM access energy is dependent on the number of accesses and the energy-per-access that varies depending upon whether the access faces \textit{a row buffer hit}, \textit{a row buffer miss}, or \textit{a row buffer conflict}.
In a row buffer hit, the requested data entity is already available in the row buffer. 
Hence, the data can be accessed directly without additional operation. 
On the other hand, a row buffer miss or conflict needs to open the requested row first before the data can be accessed.
Therefore, a row buffer hit incurs less energy and latency as compared to a row buffer miss and a row buffer conflict, as shown in Fig. \ref{Fig:BreakdownLatencyEnergyDRAM}(b)-(c).   
From these observations, it is evident that \textit{DRAM access energy for CNN accelerators can be optimized by decreasing the number of DRAM accesses, row buffer conflicts and row buffer misses}. Optimizing these parameters also lead to improved DRAM latency (and throughput) as the DRAM latency-per-access is also dependent majorly on the same parameters. 

\vspace{-0.4cm}
\subsection{The State-of-the-Art and their Limitations}
\label{Sec:SoA_Summary}

\textbf{Layer partitioning and Scheduling:} The full CNN processing usually cannot be mapped at once on the accelerator fabric due to limited on-chip memory, i.e., $100$KB-$500$KB \cite{Ref:Sze_DNNsurvey_IEEE17}. 
Therefore, \textit{layer partitioning and scheduling} are required to access a portion of data from the off-chip DRAM to the on-chip accelerator fabric for computation. 
To reduce the DRAM accesses, previous works \cite{Ref:Zhang_CNNfpga_FPGA15}\cite{Ref:Li_SmartShuttle_DATE18} employ different layer partitioning and scheduling schemes to move the data from DRAM to on-chip memory, and then reuse it 
multiple times for computation. 
Depending upon which data type\footnote{There are three data types: (1) input activations/feature maps (\textit{ifmaps}), (2) output activations/feature maps (\textit{ofmaps}), and (3) \textit{weights}.} should be kept longer on chip, previous works can be loosely classified in two categories: \textit{fixed scheduling} and \textit{adaptive scheduling}, which will be discussed in detail in Section \ref{Sec:LayerPartitionSchedule}.
In the fixed scheduling, priority of data reuse is fixed to only one data type, which is not effective to minimize the DRAM accesses, since different layers of a network may have different priority of data reuse.
This limitation is addressed by the adaptive scheduling \cite{Ref:Li_SmartShuttle_DATE18}, the state-of-the-art scheduling which adaptively prioritizes the data reuse for each layer of a network.
However, it does not consider the DRAM organization and overlapping data which should not be re-fetched again from DRAM in the analytical model for estimating the number of DRAM accesses. 
Therefore, the analytical model used for estimating the number of DRAM accesses is sub-optimal and has to be re-formulated.

\textbf{Effective DRAM Throughput:} \add{CNN accelerators typically employ DRAM \textit{burst mode} to increase the effective DRAM throughput \cite{Ref:Zhang_Caffeine_TCAD19, Ref:Qiu_GoDeep_FPGA16, Ref:Guan_FPDNN_FCCM17}, since it allows accessing multiple data with a single DRAM request}. 
The state-of-the-art work \cite{Ref:Zhang_Caffeine_TCAD19} exploits the burst mode by mapping each data partition on continuous addresses in a DRAM bank.
It prioritizes to map each data partition to different columns in the same row of the same bank. 
If all columns in the same row are fully filled, the remaining data are mapped in a different row of the same bank.
Therefore, each data partition may occupy multiple rows in a bank. 
However, this mapping has a high chance to face row buffer conflicts, since data from multiple rows in a bank need to be fetched for accessing the entire data partition, thereby consuming high energy and latency. 
Therefore, the state-of-the-art work does not optimize the DRAM data mapping, which is crucial for CNN accelerators.

\textbf{Required:} These limitations hinder the CNN accelerators to achieve further energy-efficiency improvements. 
Therefore, there is a significant need for a methodology that optimizes (1) the number of DRAM accesses by employing effective layer partitioning and scheduling, and (2) the DRAM energy-per-access and latency-per-access by employing a DRAM mapping that effectively reduces the row buffer conflicts, while improving the DRAM throughput. 
\textit{However, determining the effective layer partitioning, scheduling, and DRAM mapping, pose different design challenges as discussed below.}

\vspace{-0.3cm}
\subsection{Associated Scientific Challenges}
\label{Sec:Challenges}

\begin{figure}[hbtp]
\vspace{-0.1cm}
\centering
\includegraphics[width=\linewidth]{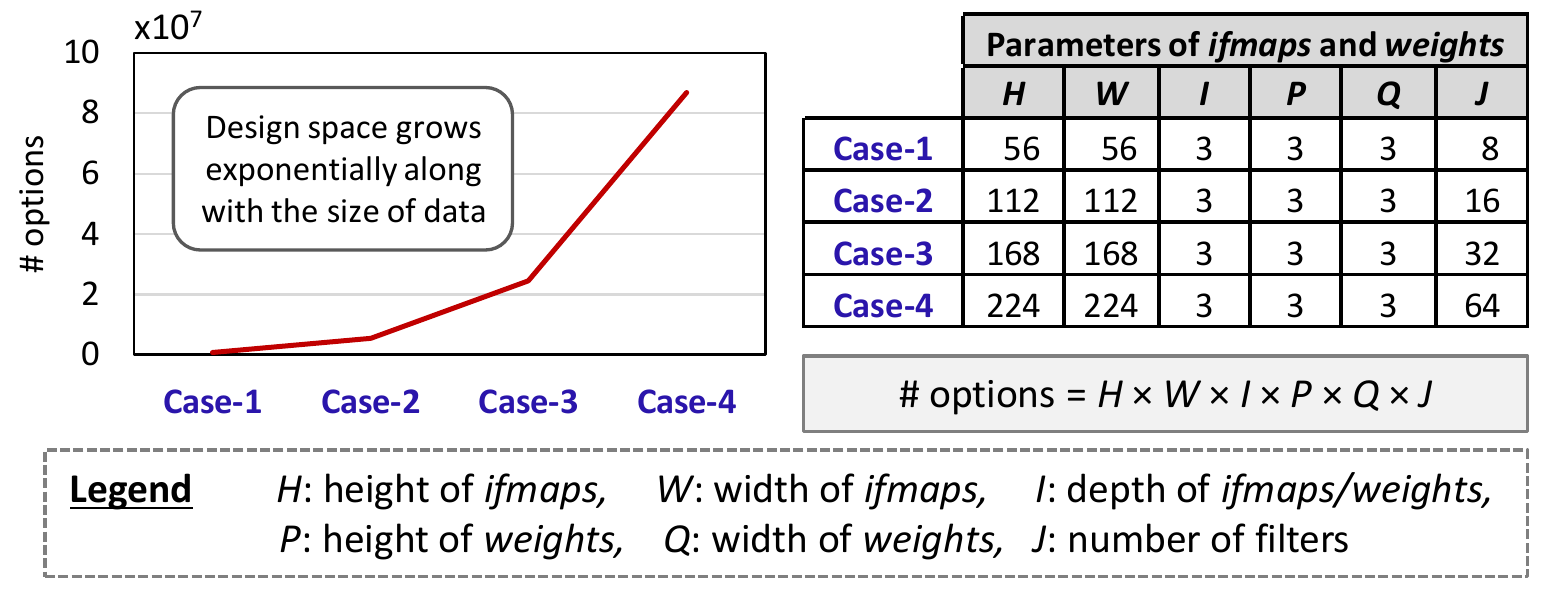}
\vspace{-0.7cm}
\caption{Estimated number of layer partitioning options to be investigated (left) in the design space for different cases listed in the table (right). Each option defines the size of \textit{ifmaps} and \textit{weights} partitions to be fetched from the DRAM and used for computing the \textit{ofmaps}.}
\label{Fig:Complexity}
\vspace{-0.4cm}
\end{figure}

%
An effective layer partitioning and scheduling solution leads to the minimum number of DRAM accesses and thereby minimum energy consumption. 
However, the number of possible configurations can be significantly large and grows exponentially with the data size, as shown in Fig.~\ref{Fig:Complexity}. Therefore, it is necessary to efficiently explore the design space in search of the effective solution/s.
Towards this, we need to formulate an optimization problem, i.e., we have to identify the constraints and formulate the design goal. 
Solving this problem requires an analytical model to efficiently estimate the number of DRAM accesses in an accurate manner. 

\vspace{-0.2cm}
\subsection{Our Novel Contributions}
\label{Sec:Contribution}

\begin{figure}[hbtp]
\vspace{-0.3cm}
\centering
\includegraphics[width=\linewidth]{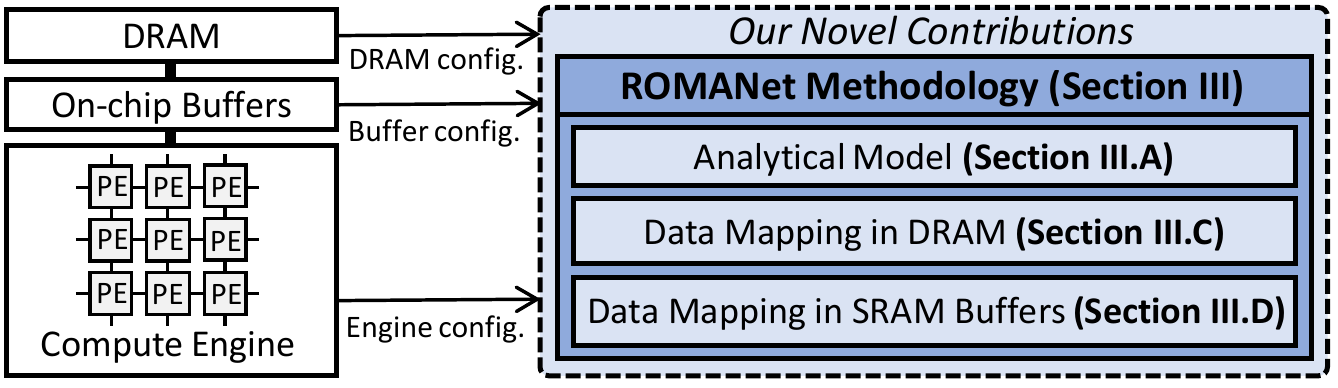}
\vspace{-0.7cm}
\caption{Typical CNN accelerator with our novel contributions in blue box.}
\label{Fig:NovelContributions}
\vspace{-0.2cm}
\end{figure}

In this paper, we make the following novel contributions (illustrated in Fig. \ref{Fig:NovelContributions}) to overcome the associated challenges. 
\begin{enumerate}[leftmargin=*]
\item We propose \textbf{ROMANet methodology} that enables fine-grained \underline{R}euse-driven \underline{O}ff-chip \underline{M}emory \underline{A}ccess management and data organization for deep neural \underline{Net}work accelerators.
\add{It minimizes (1) the DRAM accesses through a design space exploration that finds the effective layer partitioning and scheduling, and (2) the row buffer conflicts and misses through an effective DRAM mapping.} 
\item We propose \textbf{an analytical model} to compute the number of DRAM accesses in design space exploration for a given layer partitioning and scheduling of a layer. It models (1) the reuse factors of different data types (\textit{ifmaps}, \textit{ofmaps}, and \textit{weights}), and (2) the layer partitioning. 
\item We propose \textbf{a data mapping in off-chip DRAM} \add{that prioritizes the row buffer hits, the bank- and chip-level parallelism, and exploits the DRAM multi-bank burst feature. This mechanism considers the layer partitioning and DRAM configuration, and results in the throughput improvement with less access energy and latency}. 
\item We propose \textbf{a data mapping in on-chip SRAM buffers} that exploits bank-level parallelism to efficiently shuttle data between DRAM and compute engine. 
This mechanism considers the layer partitioning, buffer, and compute engine configuration. 
\end{enumerate}

\textbf{Key Results:} 
We evaluated our ROMANet using the state-of-the-art DRAM simulators \cite{Ref:Kim_Ramulator_LCA15}\cite{Ref:Ghose_VAMPIRE_POMACS18}, while considering a Tensor Processing Unit \cite{Ref:Jouppi_TPU_ISCA17}-based CNN accelerator.
The experimental results show that, compared to the state-of-the-art technique \cite{Ref:Li_SmartShuttle_DATE18}, our ROMANet reduces the DRAM access energy by 12\% for the AlexNet, by 36\% for the VGG-16, and by 46\% for the MobileNet, while improving the DRAM throughput by 10\%. 

\vspace{-0.3cm}
\subsection{Paper Organization}
\label{Sec:PaperOrg}

Section \ref{Sec:Prelim} presents the preliminaries of CNNs, layer partitioning and scheduling, and DRAM fundamentals. 
Section \ref{Sec:ROMANet} presents our ROMANet methodology. 
In Section \ref{Sec:EvalMethod}, we discuss the evaluation methodology. 
Afterwards, we discuss the experimental results and our improvements over the state-of-the-art in Section \ref{Sec:Results}.
Section \ref{Sec:Conclusion} concludes the paper.

\section{Preliminaries}
\label{Sec:Prelim}

\subsection{Convolutional Neural Networks}
\label{Sec:CNNs}

\begin{figure}[hbtp]
\vspace{-0.3cm}
\centering
\includegraphics[width=\linewidth]{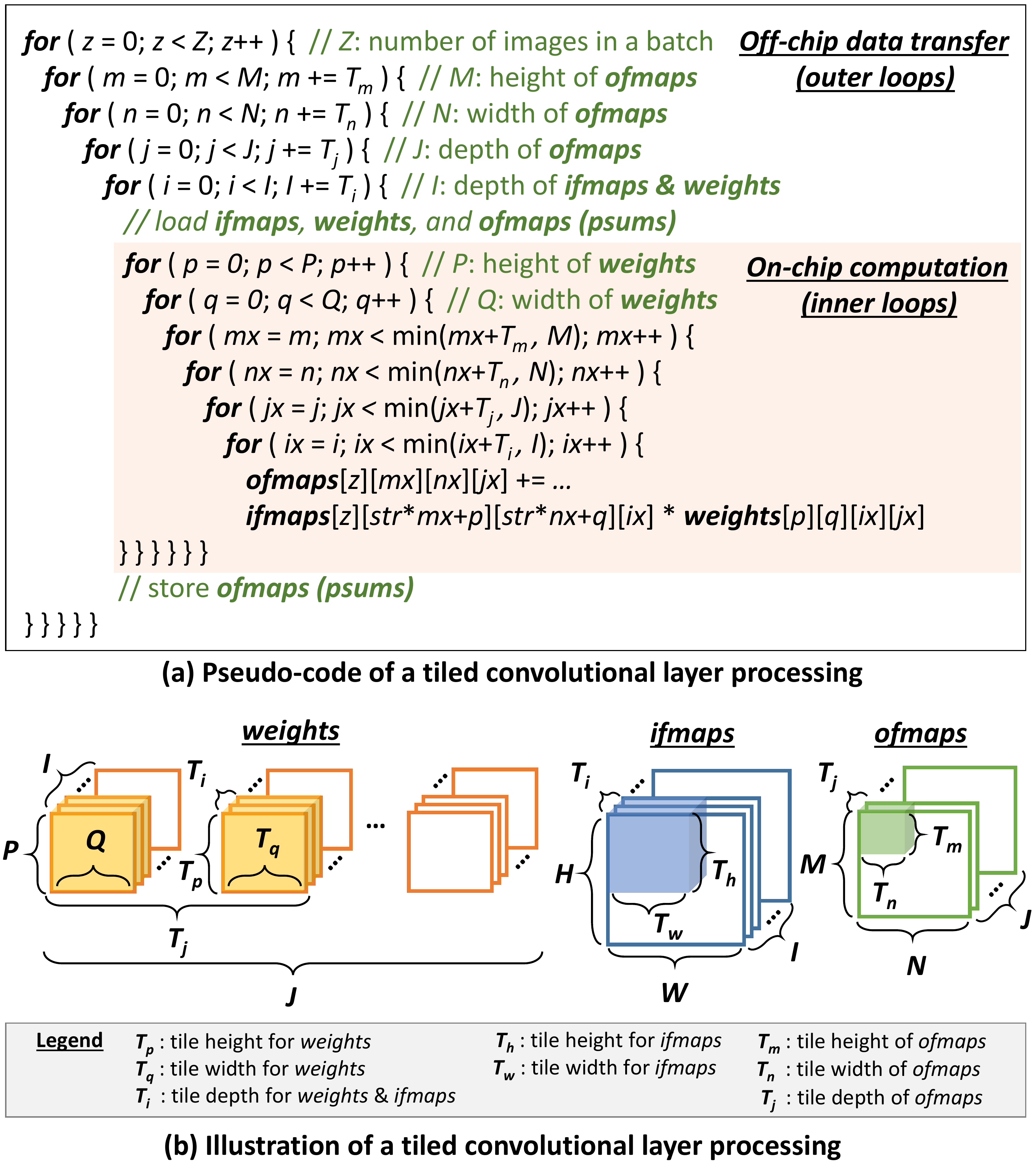}
\vspace{-0.7cm}
\caption{(a) Pseudo-code of a tile-based convolutional layer processing. (b) Illustration of a tile-based convolutional layer processing. \textit{Tiling factors} determine portion of data (shaded region) in each data type which are being considered for on-chip processing at one time.}
\label{Fig:PseudoCode_CNN}
\vspace{-0.4cm}
\end{figure}

We focus on the convolutional (CONV) and fully-connected (FC) layers which are extensively used in CNNs.
The pseudo-code of the convolutional operations is presented in Fig. \ref{Fig:PseudoCode_CNN}. 
\textit{P} and \textit{Q} denote the height and the width of \textit{weights}; \textit{M} and \textit{N} denote the height and the width of \textit{ofmaps}; \textit{I} and \textit{J} denote the number of \textit{ifmaps} and \textit{ofmaps}; \textit{H} and \textit{W} denote the height and the width of \textit{ifmaps}; and \textit{str} denotes stride of convolution, respectively.
The fully-connected operations can also be represented using the same computational loop, but by specifying the size of \textit{P}, \textit{Q}, \textit{H}, \textit{W}, \textit{M}, \textit{N}, and \textit{str} equal to $1$.

\vspace{-0.2cm}
\subsection{Layer Partitioning and Scheduling}
\label{Sec:LayerPartitionSchedule}

The layer partitioning and scheduling of convolutional operations in a CNN are typically devised for each layer of a network, hence most of the CNN accelerators process a network sequentially, one layer at a time \cite{Ref:Chen_Eyeriss_JSSC16, Ref:Zhang_CambriconX_MICRO15, Ref:Albericio_Cnvlutin_ISCA16, Ref:Luo_DaDianNao_TC17}.
The pseudo-code of a convolutional layer processing in a CNN accelerator can be represented as shown in Fig.~\ref{Fig:PseudoCode_CNN}(a). It has two main parts: the inner loops and the outer loops. 
The inner loops represent the processing of a portion of the convolutional layer whose data is available in the on-chip memory. 
The outer loops define the schedule of processing different portions. 
The size of a portion that can be processed at one time depends on the data that can be stored in the on-chip memory. 
Therefore, based on the size of the on-chip memory, the data can be partitioned in the form of blocks or tiles\footnote{Tiling is used as it can improves locality when there is data reuse~\cite{Ref:Wolf_Loop_TPDS91}.}, and this is represented with the step sizes in the outer loops in Fig.~\ref{Fig:PseudoCode_CNN}(a). 
The sequence of the outer loops defines the sequence in which these tiles are moved from the DRAM to the on-chip memory to process the corresponding portions of the layer. 
This sequence also defines the total number of DRAM accesses required to process a layer, as the two subsequent on-chip computations can have some data in common that do not need re-fetching from the DRAM and thereby affects the total number of DRAM accesses required for inference.
To minimize DRAM accesses, different layer partitioning and scheduling have been studied and employed by state-of-the-art CNN accelerators. 
The idea is to keep data (that has to be reused the most) longer in the on-chip memory.
Depending upon which data type should be kept longer in the on-chip memory, the previous works can be loosely classified in two main categories: \textit{fixed scheduling} and \textit{adaptive scheduling}.

\begin{figure}[hbtp]
\vspace{-0.3cm}
\centering
\includegraphics[width=\linewidth]{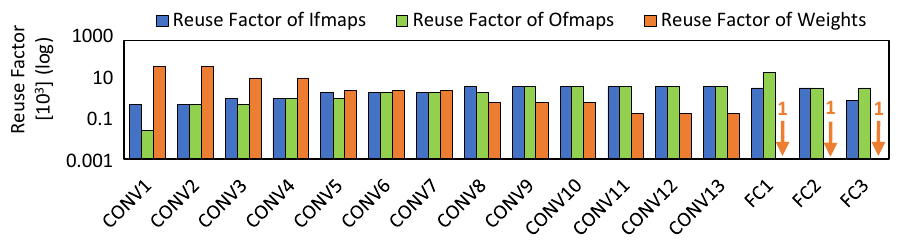}
\vspace{-0.8cm}
\caption{Reuse factor of different data types in VGG-16.}
\label{Fig:ReuseFactor_VGG16}
\vspace{-0.2cm}
\end{figure}

\textbf{Fixed scheduling} employs static tiling factors and scheduling across layers of a network, by giving priority of reuse to only one specific data type, either \textit{ifmaps}, \textit{ofmaps}, or \textit{weights}. 
This concept has been widely used in previous works, such as \cite{Ref:Zhang_CNNfpga_FPGA15}.
\textit{However, if we study the reuse factors (i.e., the number of times a specific data is reused) of different data types, they vary across layers of a CNN. Therefore, a data type which has highest reuse factor in one layer can have the least reuse factor in another layer.}
For instance, our experimental analysis in Fig. \ref{Fig:ReuseFactor_VGG16} illustrates that the \nth{1} layer of VGG-16 (CONV1) has the highest reuse factor for the \textit{weights}, however in the \nth{14} layer of VGG-16 (FC1), \textit{weights} has the lowest reuse factor.
This can lead to significant loss in energy efficiency, specifically when a fixed dataflow forces a data type, which has the highest reuse factor, to be fetched multiple times from DRAM, instead of keeping it longer in the on-chip memory and reusing it to the maximum level.

To address this limitation, \textbf{adaptive scheduling} is proposed in \cite{Ref:Li_SmartShuttle_DATE18}. 
It employs adaptive tiling factors and scheduling, and can further reduce the DRAM accesses. 
However, its analytical model for estimating the number of DRAM accesses does not consider the DRAM organization and overlapping data which should not be re-fetched from DRAM. 
Therefore, the analytical model used for estimating the number of DRAM accesses is sub-optimal and has to be re-formulated.
Furthermore, it does not consider the DRAM data organization to reduce the DRAM energy- and latency-per-access.

\vspace{-0.2cm}
\subsection{Basic DRAM Organization and Operations}
\label{Sec:DRAMorg}
\begin{figure}[hbtp]
\vspace{-0.2cm}
\centering
\includegraphics[width=\linewidth]{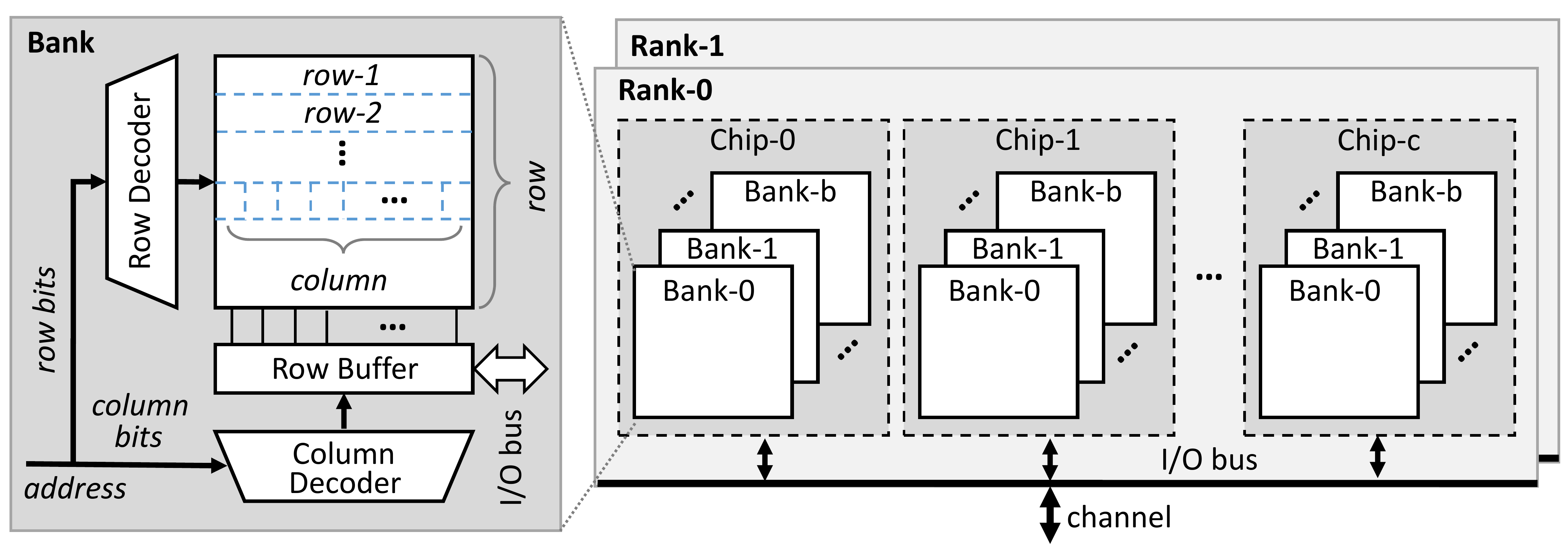}
\vspace{-0.7cm}
\caption{Overview of the DRAM organization.}
\label{Fig:DRAMorgMultiBank}
\vspace{-0.2cm}
\end{figure}

DRAM is organized into \textit{channel, rank, chip, bank, row}, and \textit{column} as seen from a top-down perspective (see Fig. \ref{Fig:DRAMorgMultiBank}). 

If there is an access request to DRAM, a specific rank will respond. 
A rank consists of a group of chips that perform all operations in lock-step \cite{Ref:Akin_DesignSpace_ASAP14} \cite{Ref:Ghose_DRAMworkload_MACS19}.
The chips in a rank are accessed in parallel and contribute forming a word. 
Inside a chip, the request is directed to a bank and decoded into row-column addresses. 
An \textit{activation} command with a row address triggers the \textit{row activation} of this bank. 
Data from a row of the DRAM cells is copied to the row buffer.
Afterwards, the \textit{read} or \textit{write} commands with a column address can be issued to the activated row buffer.
If the accessed row is already activated, then the data in this row is already in the row buffer. 
This condition is called a \textit{row buffer hit}. 
It skips the row activation step which results in reduction of DRAM access latency and energy. 
Otherwise, \textit{a row buffer miss} or \textit{a row buffer conflict} occurs, and the new referenced row needs to be activated. 
A row buffer miss means that the requested data is not available in the row buffer, since the bank does not have any row open. 
The bank needs to open the requested row first before the data can be accessed. 
Therefore, a row buffer miss requires the \textit{activation} energy to activate the target row.
A row buffer conflict means that another row is currently open and needs to be closed first before the bank can open the requested row.
To prepare the bank for an access to a different row, \textit{precharge} command is issued to trigger \textit{precharging}.
Then, the row activation step can be conducted. 
Therefore, a row buffer conflict requires \textit{precharge} and \textit{activation} energy.

\section{The ROMANet Methodology}
\label{Sec:ROMANet}

\begin{figure}[hbtp]
\centering
\includegraphics[width=\linewidth]{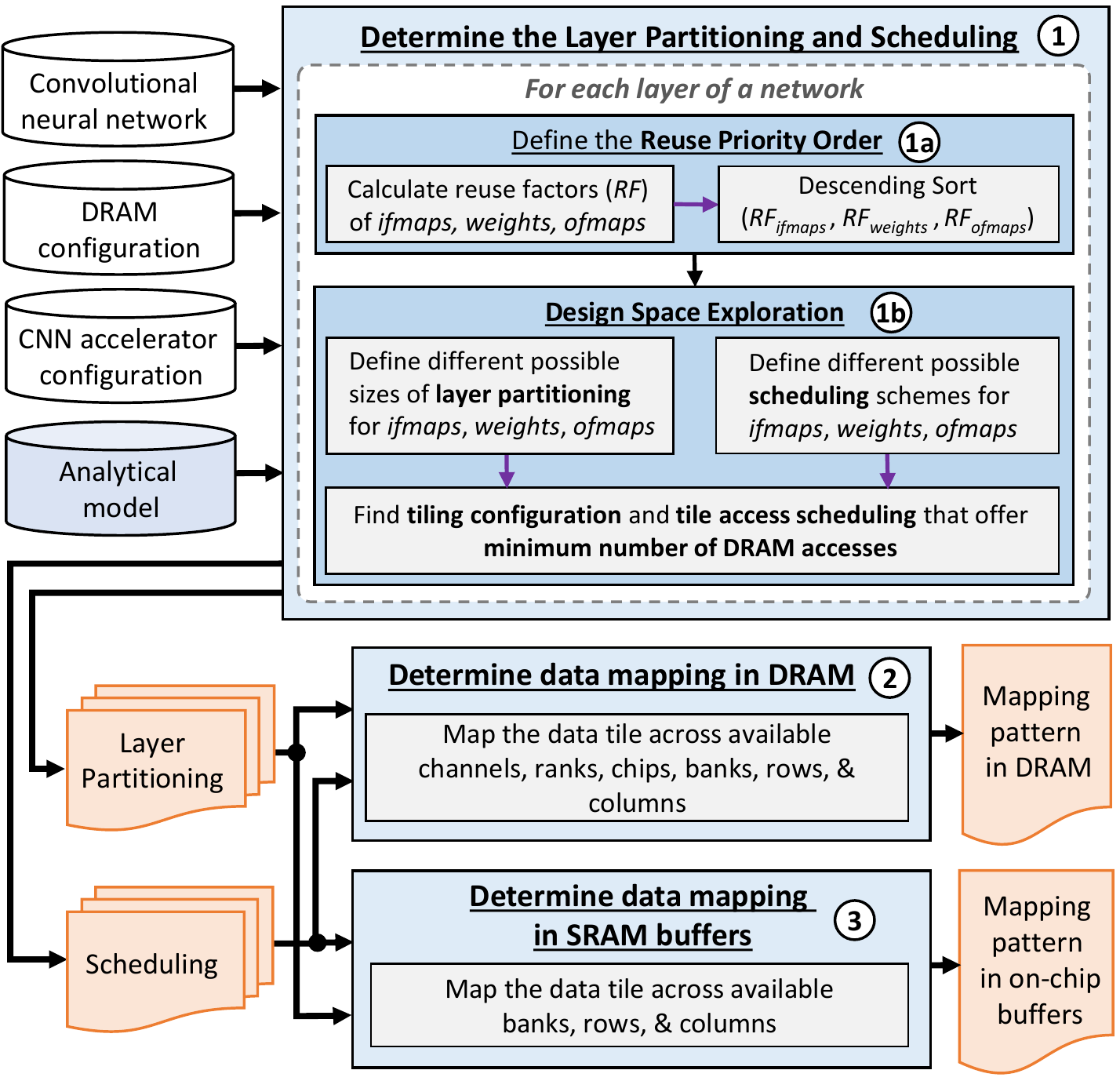}
\vspace{-0.7cm}
\caption{Operational flow of the ROMANet methodology.}
\label{Fig:ROMANetMethod}
\vspace{-0.3cm}
\end{figure}

\add{ROMANet methodology provides a novel synergistic optimization to improve the DRAM access energy and data throughput in CNN accelerators (see Fig. \ref{Fig:ROMANetMethod}).}
The overview of its operational flow is explained in the following points.
\begin{itemize}[leftmargin=*]
    \item \textbf{Step-\circled{1}:} \textbf{Determine the layer partitioning and scheduling} for each layer of a network that offer the minimum DRAM accesses. 
    However, it has to consider numerous parameters which is a non-trivial problem. 
    To solve this, \textbf{a design space exploration} (DSE) is employed. 
    The DSE requires information of the CNN (e.g., data size and stride), the DRAM (e.g., number of banks), the on-chip accelerator fabric (e.g., buffer size), and the analytical model. 
    \vspace{0.1cm}\\
    In step-\circled{1b}, the DSE considers different sizes of layer partitioning for all data types, and different scheduling schemes. 
    Towards this, we propose analytical model of layer partitioning (discussed further in \textbf{Section \ref{Sec:ROMANet_LayerPartition}}). 
    We also define the scheduling based on \textbf{the reuse priority order} from step-\circled{1a}, by analyzing the reuse factors of all data types in each layer of a network (discussed further in \textbf{Section \ref{Sec:ROMANet_ReuseFactor}}). 
    \item \textbf{Step-\circled{2}:} \textbf{The data mapping in DRAM} \add{places each data partition across the available DRAM channels, ranks, chips, banks, rows, and columns, in a manner to maximize the row buffer hits, bank- and chip-level parallelism, and to exploit the DRAM multi-bank burst feature} 
    (discussed further in \textbf{Section \ref{Sec:ROMANet_MappingDRAM}}).
    \item \textbf{Step-\circled{3}:} \textbf{The data mapping in SRAM buffers} \add{places each data partition across the available SRAM banks, rows, and columns, in a manner to maximize the bank-level parallelism} 
    (discussed further in \textbf{Section \ref{Sec:ROMANet_MappingSPM}}).
\end{itemize}

\vspace{-0.3cm}

\vspace{-0.2cm}
\subsection{Analytical Modeling for Estimating the Number of DRAM Accesses and Memory Mapping}
\label{Sec:ROMANet_AnalyticalModel}
\begin{figure}[hbtp]
\vspace{-0.2cm}
\centering
\includegraphics[width=\linewidth]{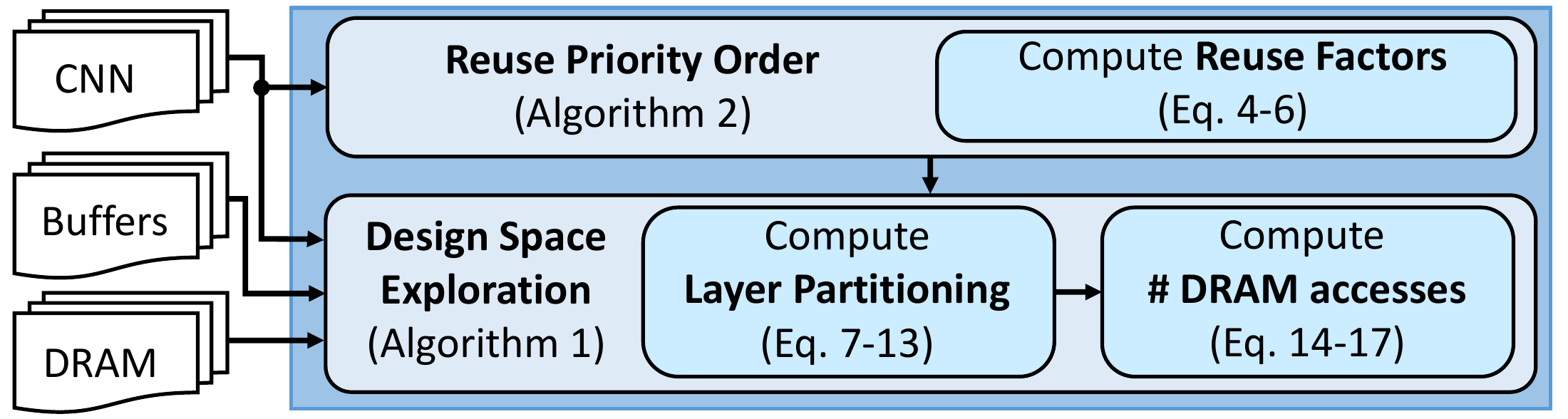}
\vspace{-0.7cm}
\caption{Flow of steps to compute the number of DRAM accesses.}
\label{Fig:EquationsFlow}
\vspace{-0.1cm}
\end{figure}

In the ROMANet, there are two key solutions proposed: (1) DSE, and (2) memory mapping in DRAM and SRAM buffers.
To perform DSE, we develop analytical model for (a) reuse factors, (b) layer partitioning, and (c) DRAM accesses. 
These models jointly provide a measure of the number of DRAM accesses that has to be minimized. 
Fig. \ref{Fig:EquationsFlow} provides an overview regarding how these models are connected. 

\textbf{Design Space Reduction:} We reduce the overall design space to be investigated by fixing some design parameters.

\begin{figure}[hbtp]
\vspace{-0.2cm}
\centering
\includegraphics[width=\linewidth]{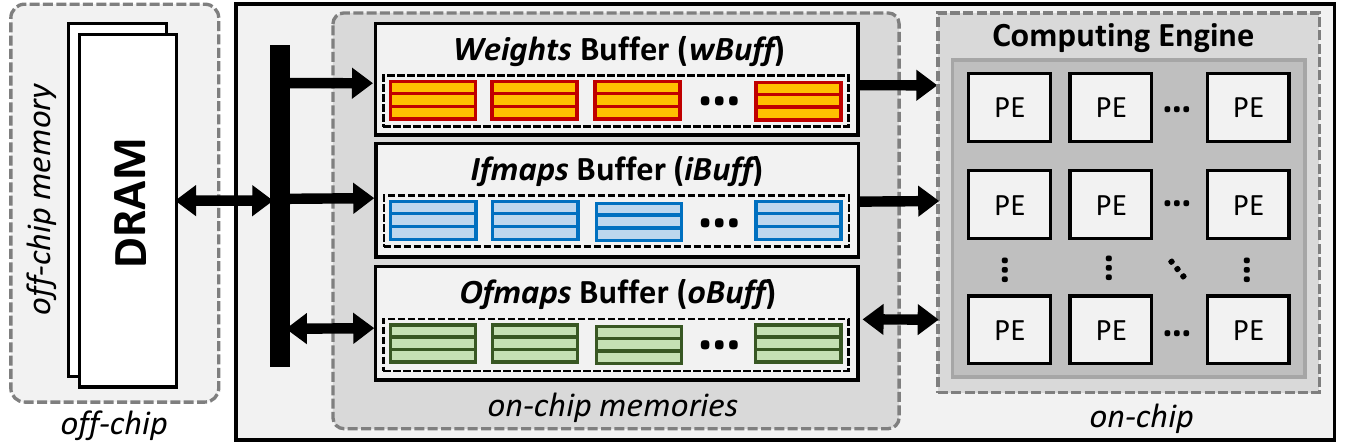}
\vspace{-0.6cm}
\caption{Typical architecture of CNN accelerator.}
\label{Fig:DNNaccelerator}
\vspace{-0.2cm}
\end{figure}

\begin{itemize}[leftmargin=*]
\item We consider a CNN accelerator design based on Tensor Processing Unit (TPU) \cite{Ref:Jouppi_TPU_ISCA17}, with a single-level SRAM buffers hierarchy implemented using \textit{scratch-pad memory} (SPM), as shown in Fig. \ref{Fig:DNNaccelerator}.
We considers SPM, since it is commonly used as the local buffer in many CNN accelerators \cite{Ref:Stoutchinin_Scheduling_arXiv19}.
\item We process only a single layer of a network at one time.
\item We also consider to process a tile of \textit{ifmaps} and \textit{weights} on-chip, and producing a tile of \textit{ofmaps} at one time. A tile of \textit{ifmaps} is defined by \textit{tiling factors} $T_h \times T_w \times T_i$, a tile of \textit{weights} is defined by $T_p \times T_q \times T_i \times T_j$, and a tile of \textit{ofmaps} is defined by $T_m \times T_n \times T_j$, as shown in Fig. \ref{Fig:PseudoCode_CNN}(b).
\end{itemize}

\vspace{0.2cm}
\textbf{The Optimization Problem Formulation:} We formulate the optimization problem to minimize the total number of DRAM accesses for a network ($\#DRAM_{access}$).  
The $\#DRAM_{access}$ is defined as the sum of the DRAM accesses from all layers of the network, expressed as
\begin{equation}
\small
  \#DRAM_{access} = \sum_{l=1}^{L} \#DRAM_{access}^{l}
\label{Eq:DRAMaccess}
\end{equation}
The $\#DRAM_{access}^{l}$ represents the total number of DRAM accesses in layer-$l$, and $L$ represents the number of layers in the network. 
\textit{Since we consider to process only a single layer of a network at one time, the optimization of $\#DRAM_{access}$ can be approached by minimizing the number of DRAM accesses-per-layer ($\#DRAM_{access}^{l}$).} 
We consider layer partitioning approach in the optimization problem, as it is employed by many CNN accelerators, such as \cite{Ref:Chen_DianNao_ASPLOS14}\cite{Ref:Zhang_CNNfpga_FPGA15}\cite{Ref:Chen_Eyeriss_JSSC16}\cite{Ref:Lu_FlexFlow_HPCA17}.
Thus, the objective of the optimization is formulated as
\begin{equation}
\small
\begin{split}
  Object&ive: \underset{<T_h, T_w, T_i, T_j, T_m, T_n>}{minimize} \#DRAM_{access}^{l} \\ 
  Constr&aints: \\ 
  & (T_h \times T_w \times T_i) \times bit_{ifmaps} \leq iBuff \\
  & (T_p \times T_q \times T_i \times T_j) \times bit_{weights} \leq wBuff \\
  & (T_m \times T_n \times T_j) \times bit_{ofmaps} \leq oBuff
\label{Eq:OptimProblemLayer}
\end{split}
\end{equation}
Constraints of the optimization problem are the size of on-chip buffers, because they limit the volume of data that can be stored at one time.
Here, the $\#DRAM_{access}^{l}$ is defined as a sum of the number of DRAM accesses from all data types in layer-$l$, expressed as
\begin{equation}
\small
\begin{split}
  \#DRAM_{access}^{l} = & \#access_{ifmaps}^{l}+\#access_{weights}^{l}+ \\ 
  & \#access_{ofmaps}^{l}
\label{Eq:DRAMaccessLayer}
\end{split}
\end{equation}
Terms $\#access_{ifmaps}^{l}$, $\#access_{weights}^{l}$, and $\#access_{ofmaps}^{l}$ represent the number of DRAM accesses in layer-$l$ for \textit{ifmaps}, \textit{weights}, and \textit{ofmaps}, respectively. 
From the above equations, it is evident that to further define the model in a more fine-grained manner, the formulations need to consider modeling of different data types for calculating the reuse factors, the layer partitioning, and the DRAM accesses.
%
\vspace{0.2cm}
\subsubsection{Model of Reuse Factors}
\label{Sec:ROMANet_ReuseFactor}
\begin{figure}[hbtp]
\vspace{-0.2cm}
\centering
\includegraphics[width=\linewidth]{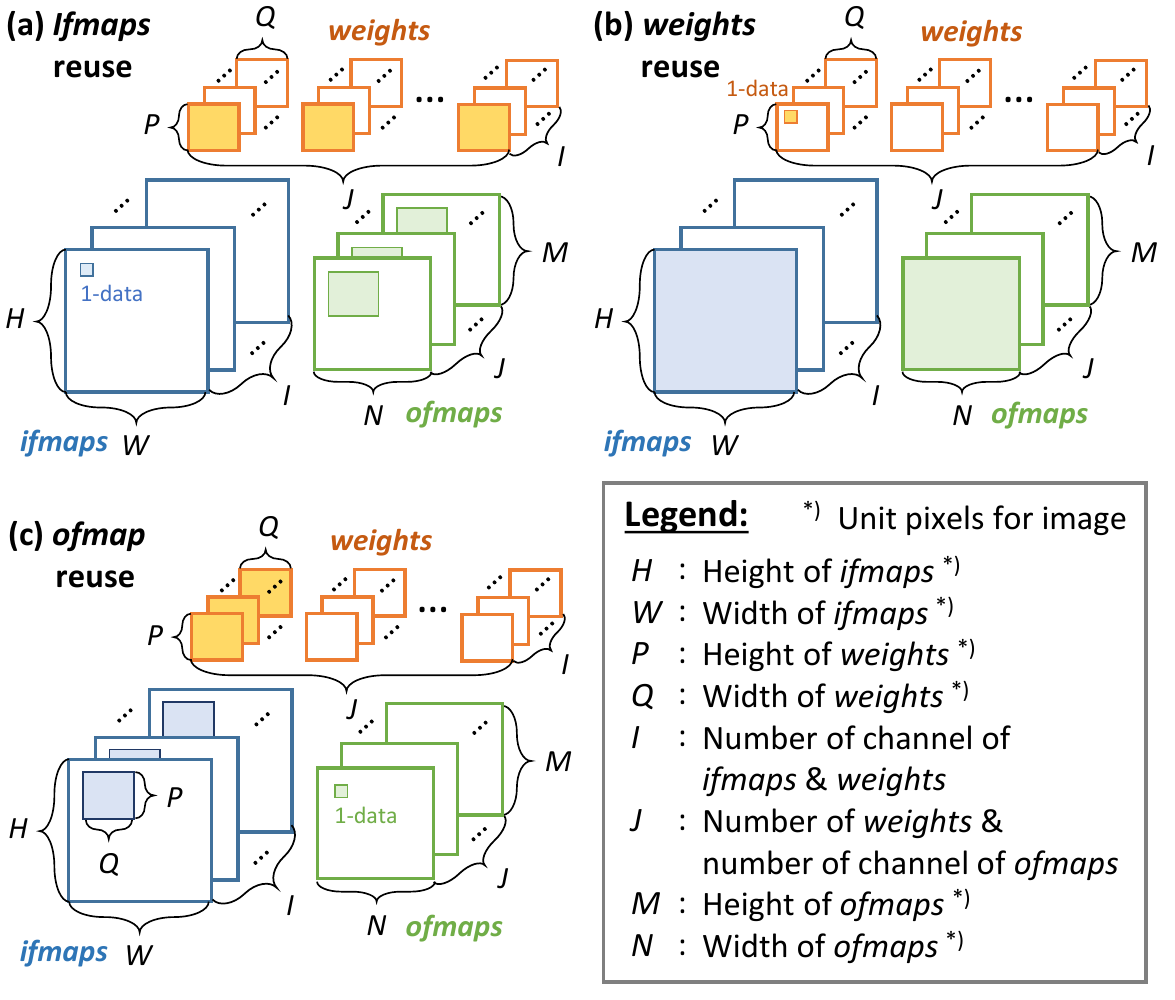}
\vspace{-0.7cm}
\caption{The reuse factor calculations in different data types involve the corresponding shaded regions. (a) In $RF_{ifmaps}$: A single pixel of \textit{ifmaps} is reused as many as the number of pixels within the shaded region in \textit{weights}. (b) In $RF_{weights}$: A single \textit{weight} is reused as many as the number of pixels within the shaded region in \textit{ifmaps}. (c) In $RF_{ofmaps}$: A single pixel of \textit{ofmaps} is reused as many as the number of \textit{partial sums} produced from convolution between the shaded regions in \textit{ifmaps} and \textit{weights}.}
\label{Fig:ReuseIllustration}
\end{figure}

Reuse factor ($RF$) represents the number of MACs that each data entity is used for \cite{Ref:Chen_EyerissV2_arXiv18}.
Hence, each data type has its own reuse factor in each layer of a network, which can be estimated by
\begin{equation}
\small
\begin{split}
  RF_{ifmaps}^{l} = \left \lceil \frac{P^{l}}{str} \right \rceil \times \left \lceil \frac{Q^{l}}{str} \right \rceil \times J^{l}
\label{Eq:RFifmap}
\end{split}
\end{equation}
\begin{equation}
\small
\begin{split}
  RF_{weights}^{l} = \left \lceil \frac {H^{l}-P^{l}+1}{str} \right \rceil \times \left \lceil \frac {W^{l}-Q^{l}+1}{str} \right \rceil
\label{Eq:RFweight}
\end{split}
\end{equation}
\begin{equation}
\small
\begin{split}
  RF_{ofmaps}^{l} = P^{l} \times Q^{l} \times I^{l}
\label{Eq:RFofmap}
\end{split}
\end{equation}
Terms $P^{l}$ and $Q^{l}$ denote the height and the width of \textit{weights}; $I^{l}$ and $J^{l}$ denote the number of \textit{ifmaps} and \textit{ofmaps}; $H^{l}$ and $W^{l}$ denote the height and the width of \textit{ifmaps}; while $str$ represents stride, respectively. 
Each of which is for a specific layer-$l$.
The \textit{reuse factor} computations for different data types are illustrated in Fig. \ref{Fig:ReuseIllustration}. 

\subsubsection{Model of Layer Partitioning}
\label{Sec:ROMANet_LayerPartition}
It is developed for all data types: (\textit{ifmaps}), \textit{weights}, and (\textit{ofmaps}). 

\begin{figure}[hbtp]
\centering
\includegraphics[width=\linewidth]{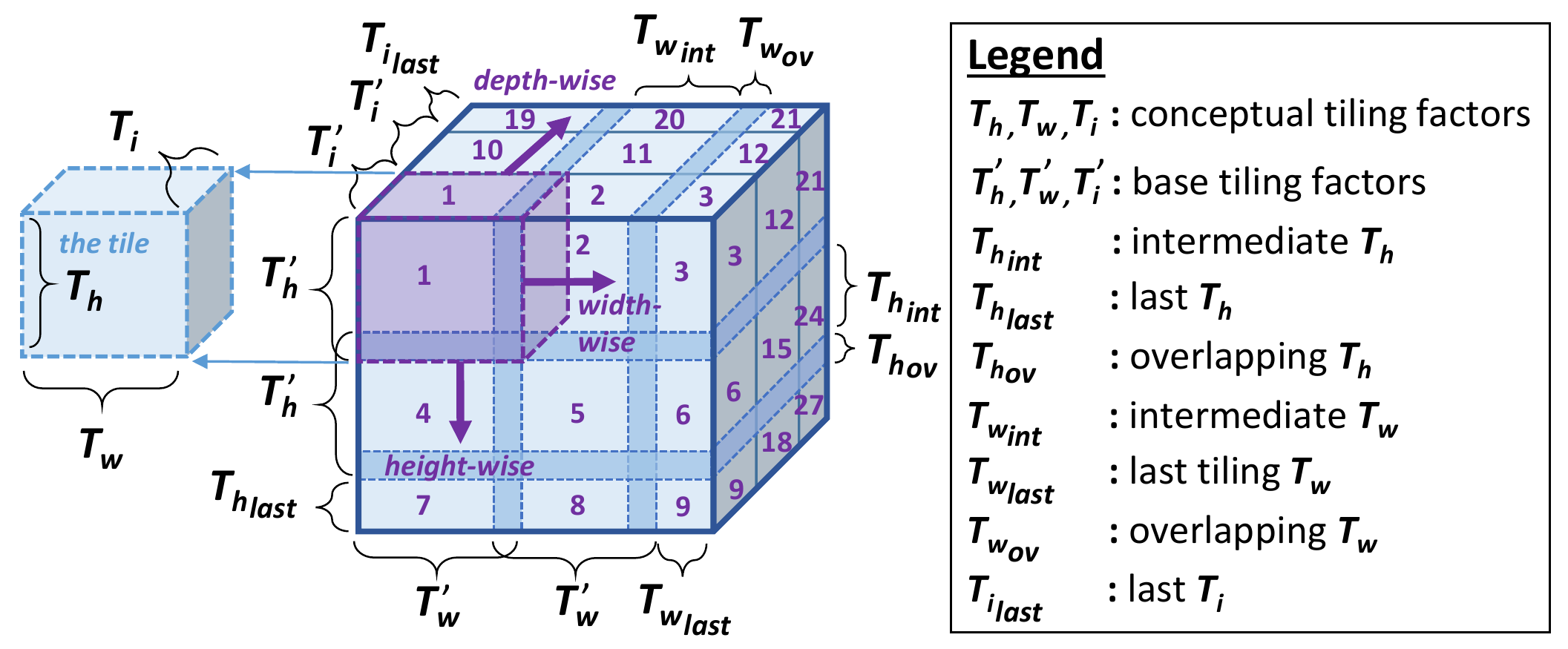}
\vspace{-0.7cm}
\caption{Model of layer partitioning for \textit{ifmaps}.} 
\label{Fig:TilingConfigPossible_Ifmap}
\vspace{-0.4cm}
\end{figure}

\begin{flushleft}
\textbf{{\textit{2.a) Input Feature Maps (ifmaps)}}}
\end{flushleft}
Depending upon the size of tiling factors, \textit{ifmaps} may have many possibilities of layer partitioning.
Therefore, multiple tiles may have different size and there may be overlapping data that do not require redundant fetches.
Here, we observe that there are three possible access directions that can affect the modeling of layer partitioning: \textit{width-wise}, \textit{height-wise}, and \textit{depth-wise}, as shown in Fig. \ref{Fig:TilingConfigPossible_Ifmap}.
%

\vspace{0.2cm} 
\underline{Width-wise direction:}
In this case, tile access scheduling for width-wise direction is prioritized. 
For instance, if we start the access from tile-1 of \textit{ifmaps}, then the sequence of access is tile-1 $\rightarrow$ tile-2 $\rightarrow$ tile-3. 
In this direction, there may be three types of tiles involved (i.e., tile-1, tile-2, and tile-3), which differ from each other in the tiling width ($T_w^{'}$: base tiling width, $T_{w_{int}}$: intermediate tiling width, and $T_{w_{last}}$: last tiling width). 
The model is formulated as
\begin{equation}
\small
\begin{split}
  T_{w_{last}} = W - T_w^{'} - n_{T_{w_{int}}}  \times \left ( T_w^{'} - T_{w_{ov}} \right )
\label{Eq:IfmapWwise2}
\end{split}
\end{equation}
Term $n_{T_{w_{int}}}$ denotes the number of intermediate tile ($T_{w_{int}}$).
The $T_{w_{int}}$ value reflects a portion of the overlapped data ($T_{w_{ov}}$) from previous fetch that should not be re-fetched again, because $T_{w_{int}} = T_w^{'} - T_{w_{ov}}$.

\vspace{0.2cm} 
\underline{Height-wise direction:}
In this case, tile access scheduling for height-wise direction is prioritized. 
For instance, if we start the access from tile-1 of \textit{ifmaps}, then the sequence of access is tile-1 $\rightarrow$ tile-4 $\rightarrow$ tile-7. 
In this direction, there may be three types of tiles involved (i.e., tile-1, tile-4, and tile-7), which differ from each other in the tiling height ($T_h^{'}$: base tiling height, $T_{h_{int}}$: intermediate tiling height, and $T_{h_{last}}$: last tiling height). 
The model is formulated as
\begin{equation}
\small
\begin{split}
  T_{h_{last}} = H - T_h^{'} - n_{T_{h_{int}}} \times \left ( T_h^{'}- T_{h_{ov}} \right )
\label{Eq:IfmapHwise2}
\end{split}
\end{equation}
Term $n_{T_{h_{int}}}$ denotes the number of intermediate tile ($T_{h_{int}}$).
The $T_{h_{int}}$ value reflects a portion of the overlapped data ($T_{h_{ov}}$) from previous fetch that should not be re-fetched again, because $T_{h_{int}} = T_h^{'} - T_{h_{ov}}$.

\vspace{0.2cm} 
\underline{Depth-wise direction:}
In this case, tile access scheduling for depth-wise direction is prioritized. 
For instance, if we start the access from tile-1 of \textit{ifmaps}, then the sequence of access is tile-1 $\rightarrow$ tile-10 $\rightarrow$ tile-19. 
In this direction, there may be two possible types of tiles involved (i.e., tile-1 and tile-19), which differ from each other in the tiling depth ($T_i^{'}$: base tiling depth and $T_{i_{last}}$: last tiling depth).
Here, there is no intermediate tile because of no overlapping data.
The model is formulated as
\begin{equation}
\small
\begin{split}
  T_{i_{last}} = I - n_{T_{i}} \times T_i^{'}
\label{Eq:IfmapIwise2}
\end{split}
\end{equation}
Term $n_{T_{i}}$ denotes the number of tile $T_{i}^{'}$.
In summary, the tiling configuration of whole \textit{ifmaps} is characterized by a combination of tiling factors $T_h \in \{T_h^{'}, T_{h_{int}}, T_{h_{last}}\}$, $T_w \in \{T_w^{'}, T_{w_{int}}, T_{w_{last}}\}$, and $T_i \in \{T_i^{'}, T_{i_{last}}\}$. 

\begin{figure}[hbtp]
\vspace{-0.2cm}
\centering
\includegraphics[width=\linewidth]{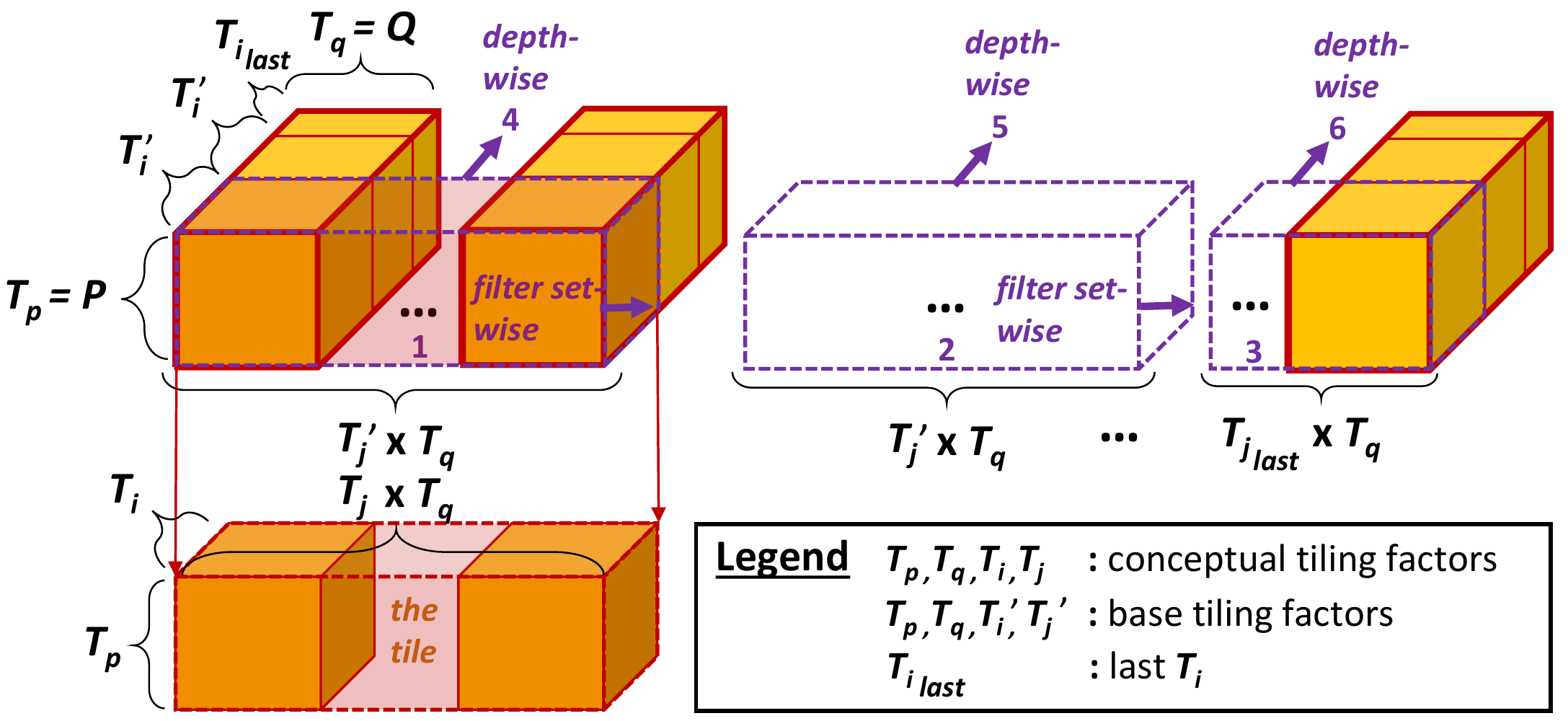}
\vspace{-0.7cm}
\caption{Model of layer partitioning for \textit{weights}.}
\label{Fig:TilingConfigPossible_Weight}
\vspace{-0.2cm}
\end{figure}

\vspace{0.2cm}
\begin{flushleft}
\textbf{\textit{2.b) Filter Weights (weights)}}
\end{flushleft}
For \textit{weights}, the layer partitioning is characterized by $T_p$, $T_q$, $T_i$, and $T_j$ as shown in Fig. \ref{Fig:TilingConfigPossible_Weight}.
Since the height and the width of \textit{weights} are typically small, we simplify the $T_p$ and $T_q$ as $T_p = P$ and $T_q = Q$ to reduce the complexity of the model and design space.
Therefore, there are only two possible access directions: \textit{filter set-wise direction} and \textit{depth-wise direction}.

\vspace{0.1cm} 
\underline{Filter set-wise direction:}
In this case, tile access scheduling for filter set-wise direction is prioritized.
The sequence of access is tile-1 $\rightarrow$ tile-2 $\rightarrow$ tile-3.
In this direction, there may be two possible types of tiles involved, which differ from each other in the number of filters in a filter-set ($T_j^{'}$: base tiling set and $T_{j_{last}}$: last tiling set).
Here, there is no intermediate tile because of no overlapping data.
The model is formulated as
\begin{equation}
\small
\begin{split}
  T_{j_{last}} = J - n_{T_{j}} \times T_j^{'}
\label{Eq:WeightJwise2}
\end{split}
\end{equation}
Term $n_{T_{j}}$ denotes the number of filter set $T_{j}^{'}$.

\vspace{0.2cm} 
\underline{Depth-wise direction:}
In this case, tile access scheduling for depth-wise direction is prioritized. 
The sequence of access is tile-1 $\rightarrow$ tile-4.
In this direction, there may be two possible types of tiles involved, which differ from each other in the tiling depth ($T_i^{'}$: base tiling depth and $T_{i_{last}}$: last tiling depth).
Here, there is no intermediate tile because of no overlapping data.
The model is formulated as
\begin{equation}
\small
\begin{split}
  T_{i_{last}} = I - n_{T_{i}} \times T_i^{'}
\label{Eq:WeightIwise2}
\end{split}
\end{equation}
Term $n_{T_{i}}$ denotes the number of tile depth $T_{i}^{'}$.
In summary, tiling configuration of \textit{weights} is characterized by a combination of tiling factors $T_p \in \{P\}$, $T_q \in \{Q\}$, $T_i \in \{T_i^{'}, T_{i_{last}}\}$, and $T_j \in \{T_j^{'}, T_{j_{last}}\}$. 

\begin{figure}[hbtp]
\vspace{-0.2cm}
\centering
\includegraphics[width=\linewidth]{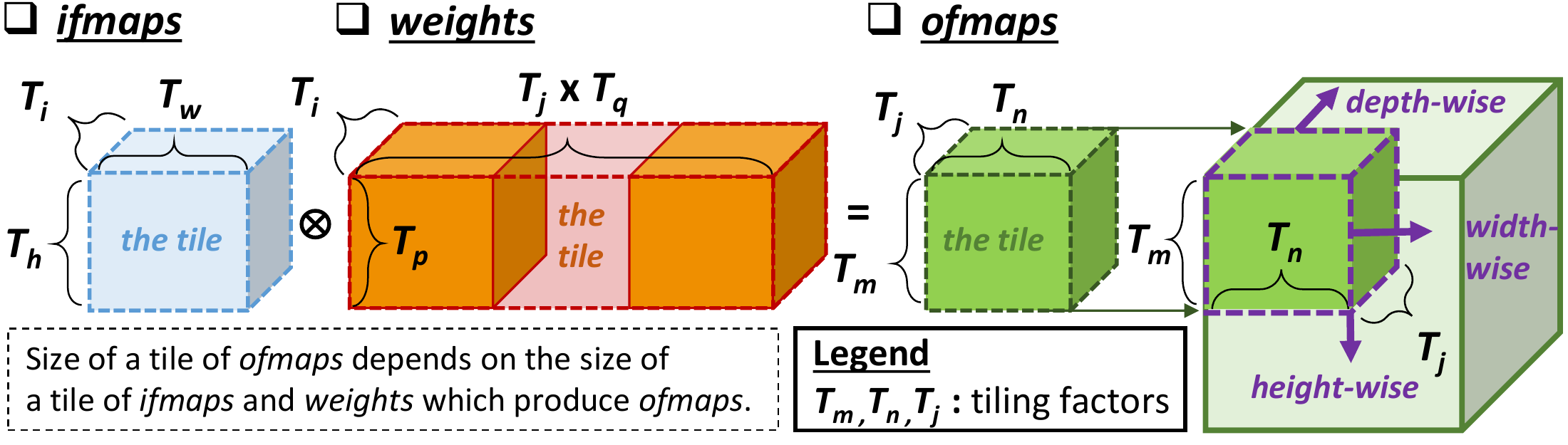}
\vspace{-0.7cm}
\caption{Model of layer partitioning for \textit{ofmaps}.}
\label{Fig:TilingConfigPossible_Ofmap}
\vspace{-0.1cm}
\end{figure}

\vspace{0.1cm}
\begin{flushleft}
\textbf{\textit{2.c) Output Feature Maps (ofmaps)}}
\end{flushleft}
For \textit{ofmaps}, the layer partitioning is characterized by $T_m$, $T_n$, and $T_j$, as illustrated in Fig. \ref{Fig:TilingConfigPossible_Ofmap}. 

A tile of \textit{ofmaps} is generated from MAC operations between a tile of \textit{ifmaps} and a tile of \textit{weights} at one time.
Which tile of \textit{ofmaps} generated from MACs, depends on which tile of \textit{ifmaps} and \textit{weights} that are computed. 
Hence, the model of \textit{ofmaps} in terms of tiling height $T_m$ and tiling width $T_n$, can be formulated as
\begin{equation}
\small
\begin{split}
  T_{m} = \left \lceil \frac{T_h - T_p + 1}{str} \right \rceil
\label{Eq:OfmapMwise}
\end{split}
\end{equation}
\begin{equation}
\small
\begin{split}
  T_{n} = \left \lceil \frac{T_w - T_q + 1}{str} \right \rceil
\label{Eq:OfmapNwise}
\end{split}
\end{equation}
Meanwhile, the model for tiling depth $T_j$ in \textit{ofmaps} follows the Eq. \ref{Eq:WeightJwise2}.
%
Even though the \textit{ifmaps} may have intermediate values such as intermediate height $T_{h_{int}}$ and intermediate width $T_{w_{int}}$, its MAC computations will still consider base tiling height $T_h^{'}$ and base tiling width $T_w^{'}$ due to overlapping data.
Therefore, in \textit{ofmaps}, the generated $T_m$ is limited to $\{T_m^{'}, T_{m_{last}}\}$ and $T_n$ is limited to $\{T_n^{'}, T_{n_{last}}\}$.
In summary, the layer partitioning of the \textit{ofmaps} is defined by combination of tiling factors $T_m \in \{T_m^{'}, T_{m_{last}}\}$, $T_n \in \{T_n^{'}, T_{n_{last}}\}$, and $T_j \in \{T_j^{'}, T_{j_{last}}\}$. 

\vspace{0.3cm}
\subsubsection{Model of DRAM Accesses}
\label{Sec:ModelMemAccess}
Eq. \ref{Eq:DRAMaccessLayer} shows that the DRAM access-per-layer is defined as the sum of the DRAM accesses from all data types.
Each of which is dependent on the layer partitioning. 
Therefore, the number of DRAM accesses-per-layer for each data type can be estimated as the sum of the number of DRAM accesses-per-tile, and can be formulated as
\begin{equation}
\small
\begin{split}
  \#access_{x}^{l} & = \sum_{t=1}^{T_{x}} \#access_{x}^{t}
  \\
  \text{with} \; \; x & \in \{ifmaps, weights, ofmaps\}
\label{Eq:DRAMaccess_IfmapPerLayer}
\end{split}
\end{equation}
Term $\#access_{x}^{t}$ represents the number of DRAM accesses-per-tile for $x$ data type ($x$ is either \textit{ifmaps}, \textit{weights}, or \textit{ofmaps}), while $T_{x}$ denotes the number of tiles in layer-$l$, for $x$ data type. 
In DRAM, \textit{ifmaps} and \textit{weights} have only read ($rd$) type of accesses, thus their number of accesses-per-tile can be estimated as
\begin{equation}
\small
\begin{split}
  \#access_{ifmaps}^{t} = & {\left \lceil \frac{T_h \times T_w \times T_i}{D_{p}} \right \rceil }_{rd}
\label{Eq:DRAMaccess_IfmapPerTile}
\end{split}
\end{equation}
\begin{equation}
\small
\begin{split}
  \#access_{weights}^{t} = & {\left \lceil \frac{T_p \times T_q \times T_i \times T_j}{D_{p}} \right \rceil }_{rd}
\label{Eq:DRAMaccess_WeightPerTile}
\end{split}
\end{equation}
Term $D_{p}$ denotes the number of DRAM chips-per-rank.
Meanwhile, the \textit{ofmaps} may have two types of DRAM accesses, read ($rd$) and write ($wr$). 
These two types exist when a tile of \textit{partial sums} which are in \textit{oBuff} can not be accumulated with newly generated \textit{partial sums}, but they still need to be accumulated with other \textit{partial sums} to produce final \textit{ofmaps}. 
Hence, they need to be stored back to DRAM, so that the \textit{oBuff} can provide space to store a new tile of \textit{partial sums}. 
Later, this tile of \textit{partial sums} has to be fetched from DRAM to complete the computation, producing a tile of final \textit{ofmaps}.
The generic equation to estimate the number of DRAM accesses-per-tile for \textit{ofmaps} is formulated as
\begin{equation}
\small
\begin{split}
  \#access_{ofmaps}^{t} = & {\left \lceil \frac{T_m \times T_n \times T_j}{D_{p}} \right \rceil }_{rd} + \\
  & {\left \lceil \frac{T_m \times T_n \times T_j}{D_{p}} \right \rceil}_{wr}
\label{Eq:DRAMaccess_OfmapPerTile}
\end{split}
\end{equation}
Eq. \ref{Eq:DRAMaccess_OfmapPerTile} shows that the same \textit{partial sums} are moved from \textit{oBuff} to DRAM ($wr$) and fetched again from DRAM to \textit{oBuff} ($rd$).
If there is no need for transporting \textit{partial sums} back to DRAM, the equation only needs to consider write ($wr$) access part for storing final \textit{ofmaps} to DRAM.

\vspace{-0.2cm}
\subsection{Proposed Design Space Exploration (DSE) for Solving the Optimization Problem}
\label{Sec:ROMANet_DSE}

\begin{figure}[hbtp]
\vspace{-0.2cm}
\centering
\scriptsize
\includegraphics[width=\linewidth]{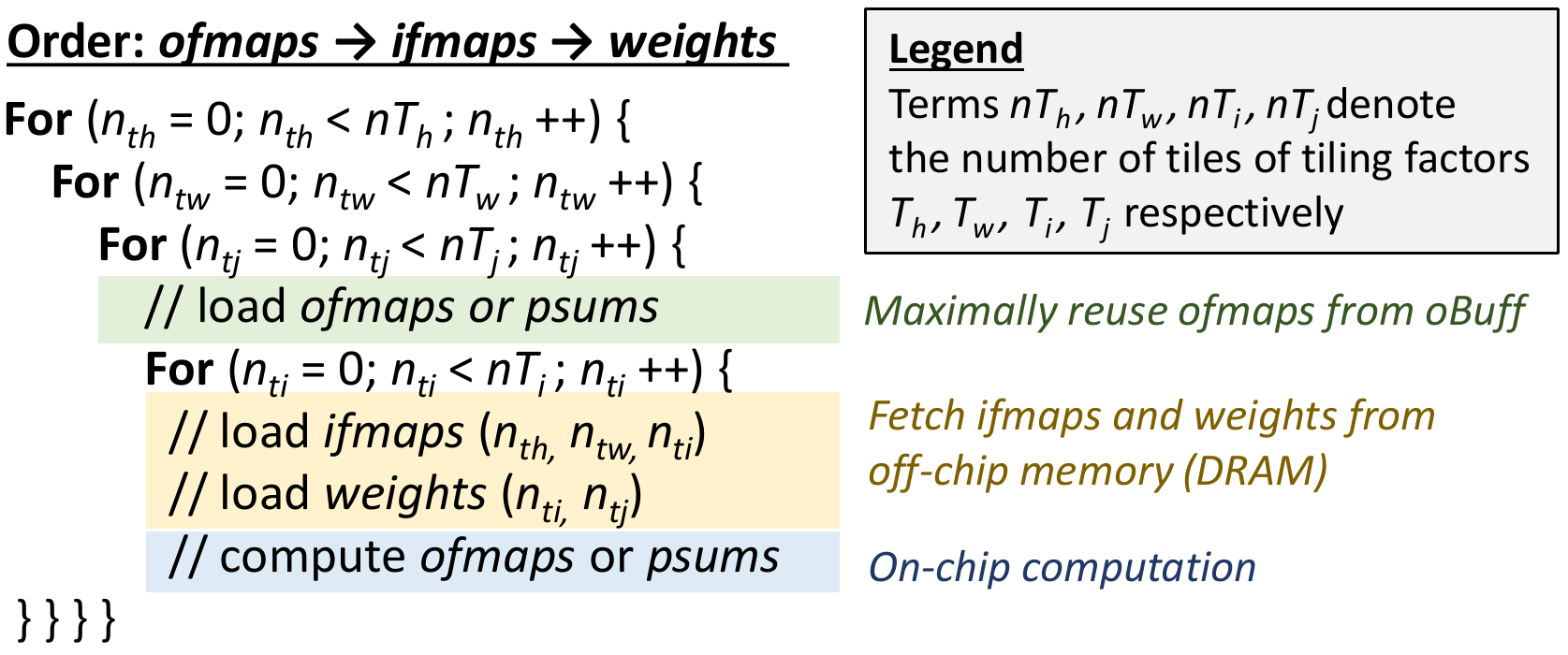}
\vspace{-0.6cm}
\caption{Illustration of a scheduling scheme derived from the reuse priority order of \textit{ofmaps} $\rightarrow$ \textit{ifmaps} $\rightarrow$ \textit{weights}. Here, \textit{ofmaps} is maximally reused by re-fetching the \textit{partial sums} from \textit{oBuff} to produce final \textit{ofmaps}. Meanwhile, the tiling parameters of \textit{ifmaps} (i.e., $nT_h$, $nT_w$) are put at the outer loop to achieve less redundant accesses than \textit{weights}.}
\label{Fig:TileAccessPattern}
\vspace{-0.2cm}
\end{figure}

\textit{We devise an algorithm that performs an exhaustive DSE for searching the effective layer partitioning and scheduling that offer minimum DRAM accesses.}
The algorithm is presented in Alg. \ref{Alg:DSE} and explained in the following steps.
%
\begin{algorithm}
 \small
 \caption{Pseudo-code of the Proposed DSE}
 \label{Alg:DSE}
 \begin{algorithmic}[1]
 \renewcommand{\algorithmicrequire}{\textbf{INPUT:}}
 \renewcommand{\algorithmicensure}{\textbf{OUTPUT:}}
 \REQUIRE \textbf{(1)} CNN configuration: number of layers ($L$), \textit{ifmaps} ($H$, $W$, $I$), \textit{weights} ($P$, $Q$, $I$), \textit{ofmaps} ($M$, $N$, $J$), etc.; \\ 
 \textbf{(2)} Buffer size: \textit{ifmaps} (\textit{iBuff}), \textit{weights} (\textit{wBuff}), \textit{ofmaps} (\textit{oBuff}); \\
 \textbf{(3)} Bitwidth: \textit{ifmaps} ($bit_{ifm}$), \textit{weights} ($bit_{wgh}$), \textit{ofmaps} ($bit_{ofm})$; \\
 \textbf{(4)} Analytical models: (i) tiling factors of \textit{ifmaps} ($T_h$, $T_w$, $T_i$), \textit{weights} ($T_p$, $T_q$, $T_i$, $T_j$), \textit{ofmaps} ($T_m$, $T_n$, $T_j$); (ii) DRAM accesses;\\
 \textbf{(5)} Reuse Priority Orders ($RPO$); // from Alg. \ref{Alg:ReusePriorityOrder} \\
 \ENSURE \textbf{(1)} Number of DRAM accesses ($\#DR_{access}$); \\
 \textbf{(2)} Layer partitioning: \textit{ifmaps} ($TP_{ifm}$), \textit{weights} ($TP_{wgh}$), \textit{ofmaps} ($TP_{ofm}$); \\
 \textbf{(3)} Scheduling ($Schedule$); \\
 \vspace{0.1cm}
\renewcommand{\algorithmicrequire}{\textbf{BEGIN}}
\renewcommand{\algorithmicensure}{\textbf{END}}
\REQUIRE \hspace{0.1cm} \\ 
 \textbf{Initialization}: \\
 \STATE $T_p = P$; \\
 \STATE $T_q = Q$; \\
 \textbf{Process}: \\
 \STATE $\#Scheduling$ $\leftarrow$ $RPO$;
 \FOR{($Layer = 1$ to $L$)}
   \FOR{($Sched = 1$ to $\#Scheduling$)}
     \FOR{($T_h = P : step_{T_h} : H$)}
       \FOR{($T_w = Q : step_{T_w} : W$)}
         \FOR{($T_j = 1 : step_{T_j} : J$)}
           \STATE Calculate $T_i$;\\ 
           \IF{($T_h \times T_w \times T_i \times bit_{ifm} \leq iBuff$) \AND \\ 
           ($P \times Q \times T_i \times T_j \times bit_{wgh} \leq wBuff$)}
               \STATE Calculate $T_m$, $T_n$, $T_j$; \\
               \IF{($T_m \times T_n \times T_j \times bit_{ofm} \leq oBuff$)}
                 \STATE Calculate $\#DR_{access}$;\\
                 \IF{(first loop)}
                   \STATE $minDR_{access} = \#DR_{access}$;\\
                   \STATE Save $TP_{ifm}$, $TP_{wgh}$, and $TP_{ofm}$;\\
                   \STATE Save $Schedule = Sched$;\\
                 \ELSIF{($\#DR_{access} \leq minDR_{access}$)}
                   \STATE $minDR_{access} = \#DR_{access}$;\\
                   \STATE Save $TP_{ifm}$, $TP_{wgh}$, and $TP_{ofm}$;\\
                   \STATE Save $Schedule = Sched$;\\
                 \ENDIF\\
               \ENDIF\\
           \ENDIF\\
         \ENDFOR\\
       \ENDFOR\\
     \ENDFOR\\
   \ENDFOR\\
 \ENDFOR\\
 \RETURN (1) $minDR_{access}$, (2) $TP_{ifm}$, $TP_{wgh}$, $TP_{ofm}$, \\ (3) $Schedule$; \\
 \ENSURE
 \end{algorithmic}
 \end{algorithm}
\begin{algorithm}
 \small
 \caption{Pseudo-code of the Reuse Priority Order}
 \label{Alg:ReusePriorityOrder}
 \begin{algorithmic}[1]
 \renewcommand{\algorithmicrequire}{\textbf{INPUT:}}
 \renewcommand{\algorithmicensure}{\textbf{OUTPUT:}}
 \REQUIRE \textbf{(1)} CNN configuration: number of layers ($L$), \textit{ifmaps} ($H$, $W$, $I$), \textit{weights} ($P$, $Q$, $I$), \textit{ofmaps} ($M$, $N$, $J$); \\ 
 \textbf{(2)} Analytical model: reuse factor ($RF$); 
 \ENSURE  Reuse Priority Orders ($RPO$); // for Alg. \ref{Alg:DSE}\\
 \vspace{0.1cm}
 \renewcommand{\algorithmicrequire}{\textbf{BEGIN:}}
 \renewcommand{\algorithmicensure}{\textbf{END:}}
 \REQUIRE \hspace{0.1cm} \\ 
 \textbf{Initialization}: \\
 \STATE $RF_{ifm}, RF_{wgh}, RF_{ofm}  = 0$; \\
 \STATE $RPO  = []$; \\
 \textbf{Process}: \\
 \FOR{($Layer = 1$ to $L$)}
   \STATE Calculate $RF_{ifm}$, $RF_{wgh}$, $RF_{ofm}$; \\ 
   \STATE $RPO[Layer] \leftarrow Sort (RF_{ifm}, RF_{wgh}, RF_{ofm})$; \\
 \ENDFOR\\
 \RETURN $RPO$; \\
 \ENSURE
 \end{algorithmic} 
 \end{algorithm}
 \setlength{\textfloatsep}{6pt}
%
\vspace{0.2cm}\\
\textbf{(Step-1)} 
For each layer of a network, we define scheduling schemes to be explored in the DSE, which are based on the reuse priority orders that are determined using Alg. \ref{Alg:ReusePriorityOrder}.
\vspace{0.2cm}\\
In Alg. \ref{Alg:ReusePriorityOrder}, the reuse factors of different data types are calculated and then sorted, so that the order of priority is known (Alg. \ref{Alg:ReusePriorityOrder}: lines 3-5). 
The idea is to ensure that the data type with a higher reuse factor has a higher priority to be kept longer in the on-chip buffer and reused maximally for reducing the redundant accesses to DRAM.
The possible orders are presented in Table \ref{Table:DataReuseStrategy}.
\begin{table}[hbtp]
\vspace{-0.2cm}
\caption{Possible reuse priority orders for scheduling.}
\vspace{-0.1cm}
\centering
\begin{tabular}{|c|c|c|c|} 
\hline
\multirow{2}{*}{\textbf{Order}} & \multicolumn{3}{c|}{\textbf{Reuse Factors}} \\
\cline{2-4}
& \textbf{Highest} & \textbf{Medium} & \textbf{Lowest}  \\
\hline 
\hline
1 & \textit{ifmaps} & \textit{weights} & \textit{ofmaps}
\\
\hline
2 & \textit{ifmaps} & \textit{ofmaps} & \textit{weights} 
\\
\hline
3 & \textit{weights} & \textit{ifmaps} & \textit{ofmaps}
\\
\hline
4 & \textit{weights} & \textit{ofmaps} & \textit{ifmaps}
\\
\hline
5 & \textit{ofmaps} & \textit{ifmaps} & \textit{weights}
\\
\hline
6 & \textit{ofmaps} & \textit{weights} & \textit{ifmaps}    
\\
\hline
\end{tabular}
\label{Table:DataReuseStrategy}
\vspace{-0.1cm}
\end{table}
\vspace{0.2cm}\\
Therefore, a network will have reuse priority orders from all layers which can be used to define scheduling schemes for DSE (Alg. \ref{Alg:DSE}: line 3). 
For instance, if a layer has reuse priority order \textit{ofmaps} $\rightarrow$ \textit{ifmaps} $\rightarrow$ \textit{weights}, then the corresponding scheduling is devised as shown in Fig. \ref{Fig:TileAccessPattern}.
Here, the priority is to maximally reuse \textit{ofmaps} or partial sums (\textit{psums}) by traversing the tile accesses of \textit{ifmaps} and \textit{weights} in the depth-wise direction.
Therefore, the DRAM accesses for \textit{ofmaps} are minimized, i.e., only for storing the final \textit{ofmaps} to the DRAM.
Furthermore, \textit{ifmaps} should have less redundant accesses than \textit{weights}, since \textit{ifmaps} has higher reuse priority than \textit{weights}.
It is expected to be achieved by putting the tiling parameters of \textit{ifmaps} (i.e., $nT_h$, $nT_w$) at the outer loop of the scheduling.
%
%
\vspace{0.2cm}\\
\textbf{(Step-2)} 
Fetch a tile of \textit{ifmaps} and \textit{weights} from DRAM, as long as they fit in the corresponding buffer (\textit{iBuff} and \textit{wBuff}, respectively) and can be used together in MAC operations for generating a tile of \textit{ofmaps}, whose size has to fit in the \textit{oBuff} (Alg. \ref{Alg:DSE}: lines 10-12).
To define the size of a tile, we explore different combinations of tiling factors from \textit{ifmaps}, \textit{weights}, and \textit{ofmaps} (Alg. \ref{Alg:DSE}: lines 6-12).
Here, \textit{we consider the adjustable search steps for different tiling factors, i.e., $step_y$, with $y \in \{T_h, T_w, T_j\}$, to achieve faster search, as the higher search step means smaller design space to be explored.}
This gives trade-offs between the possible number of DRAM accesses that can be found and the DSE computation time.
\vspace{0.2cm}\\
\textbf{(Step-3)}
DSE calculates the number of DRAM accesses (in Alg. \ref{Alg:DSE}: line 13) based on the layer partitioning and scheduling, which have been defined in the previous steps.
\vspace{0.2cm}\\
\textbf{(Step-4)}
The information of layer partitioning and scheduling that offer minimum DRAM accesses is saved (Alg. \ref{Alg:DSE}: lines 14-22), and then used for mapping data in DRAM and buffers, which will be discussed in Section \ref{Sec:ROMANet_MappingDRAM} and Section \ref{Sec:ROMANet_MappingSPM}, respectively.

\vspace{0.2cm}
\add{\textbf{Note}: The DSE needs to be performed only once at the design time to find the effective partitioning and scheduling policy.  
Once the policy has been found, the corresponding settings in the CNN accelerator are set once during the initialization stage before performing an inference.} 

\vspace{-0.3cm}
\subsection{Data Mapping in the DRAM}
\label{Sec:ROMANet_MappingDRAM}

As the dataflow of CNN processing is known prior to the execution, it is always certain at every point during the execution that which data will be required next. 
Therefore, CNN execution can significantly benefit from the spatial locality. 
Spatial locality means that if a particular data is referenced at a particular time, it is likely that the adjacent data will be referenced in the near future. 
In CNN execution where the dataflow is completely known, \textit{the spatial locality can be maximized by mapping the data that will be required in the subsequent cycles together in the DRAM.}

\begin{figure}[hbtp]
\vspace{-0.2cm}
\centering
\includegraphics[width=\linewidth]{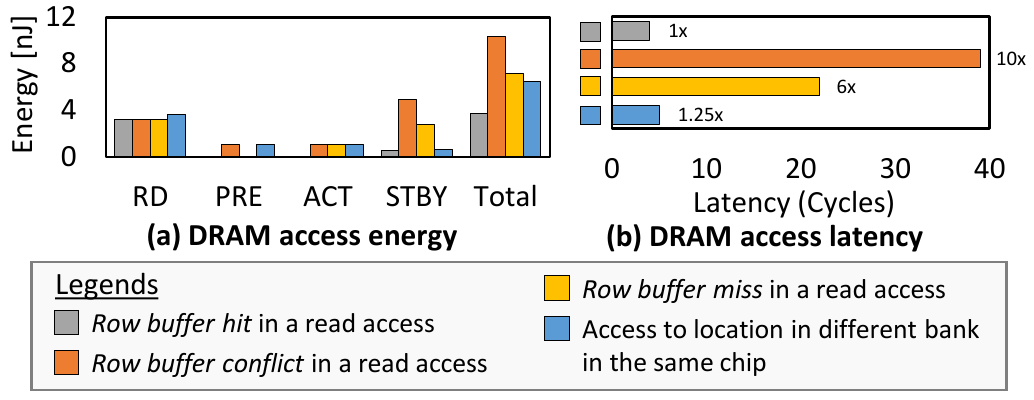}
\vspace{-0.7cm}
\caption{(a) The DRAM access energy incurred in different conditions, and \newline (b) the DRAM access latency incurred in different conditions. 
Data are obtained from our experiments using state-of-the-art cycle-accurate DRAM simulators \cite{Ref:Kim_Ramulator_LCA15} \cite{Ref:Ghose_VAMPIRE_POMACS18} for DDR3-1600 2Gb x8.}
\label{Fig:DDR3_EnergyLatency}
\vspace{-0.1cm}
\end{figure}

\begin{figure}[hbtp]
\vspace{-0.3cm}
\centering
\includegraphics[width=\linewidth]{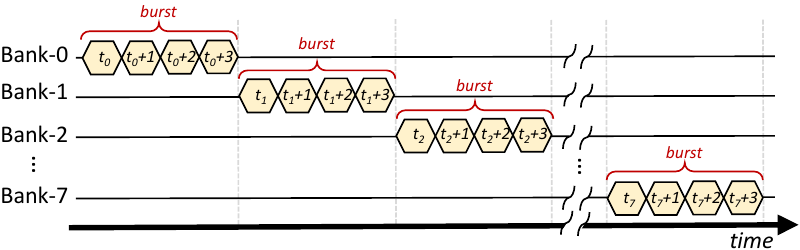}
\vspace{-0.7cm}
\caption{\add{Timing diagram of DRAM multi-bank burst feature, which happen in the same DRAM chip.}}
\label{Fig:DRAM_MultiBankBurst}
\vspace{-0.2cm}
\end{figure}

To understand the possible energy and latency values in a single DRAM access, we performed experiments to observe the energy and latency incurred by a single access in different conditions: \textit{a row buffer hit}, \textit{a row buffer conflict}, \textit{a row buffer miss}, and an access to location in different bank in the same chip.
The experimental results for energy and latency are presented in Fig.~\ref{Fig:DDR3_EnergyLatency}(a) and Fig.~\ref{Fig:DDR3_EnergyLatency}(b), respectively.
\\
These figures show some key observations:
\begin{itemize}[leftmargin=*]
    \item \addx{A single DRAM access consumes \textit{standby} (STBY) energy and operational energy: \textit{read} (RD) or \textit{write} (WR), \textit{activation} (ACT), and \textit{precharge} (PRE) \cite{Ref:Ghose_DRAMworkload_MACS19}.}
    \item A row buffer hit requires \textit{read}/\textit{write} and \textit{standby} energy, thereby consuming less access energy and latency compared to a row buffer conflict or a row buffer miss.
    \item A row buffer conflict requires \textit{precharge}, \textit{activation}, and \textit{standby} energy, as it has to close the currently activated row and then open a different row. It consumes the highest access energy and latency among the observed cases.
    \item A row buffer miss requires \textit{activation} and \textit{standby} energy to activate the target row as there is no activated row yet.
    \item Exploiting DRAM bank-level parallelism (i.e., access to location in different bank of the same chip in a short time interval) can also be done faster and consume less energy compared to a row buffer conflict. \addx{It requires the \textit{activation}, \textit{precharge} and \textit{standby} energy, but with less total energy than a row buffer conflict, due to its lower latency}. 
\end{itemize}
\add{Furthermore, DRAM has the multi-bank burst feature that can be exploited to further improve the data throughput, as shown in Fig.~\ref{Fig:DRAM_MultiBankBurst}.
Therefore, we also devise the effective DRAM mapping that offers high throughput with the lowest access energy and latency.} 

\begin{figure}[hbtp]
\vspace{-0.3cm}
\centering
\includegraphics[width=\linewidth]{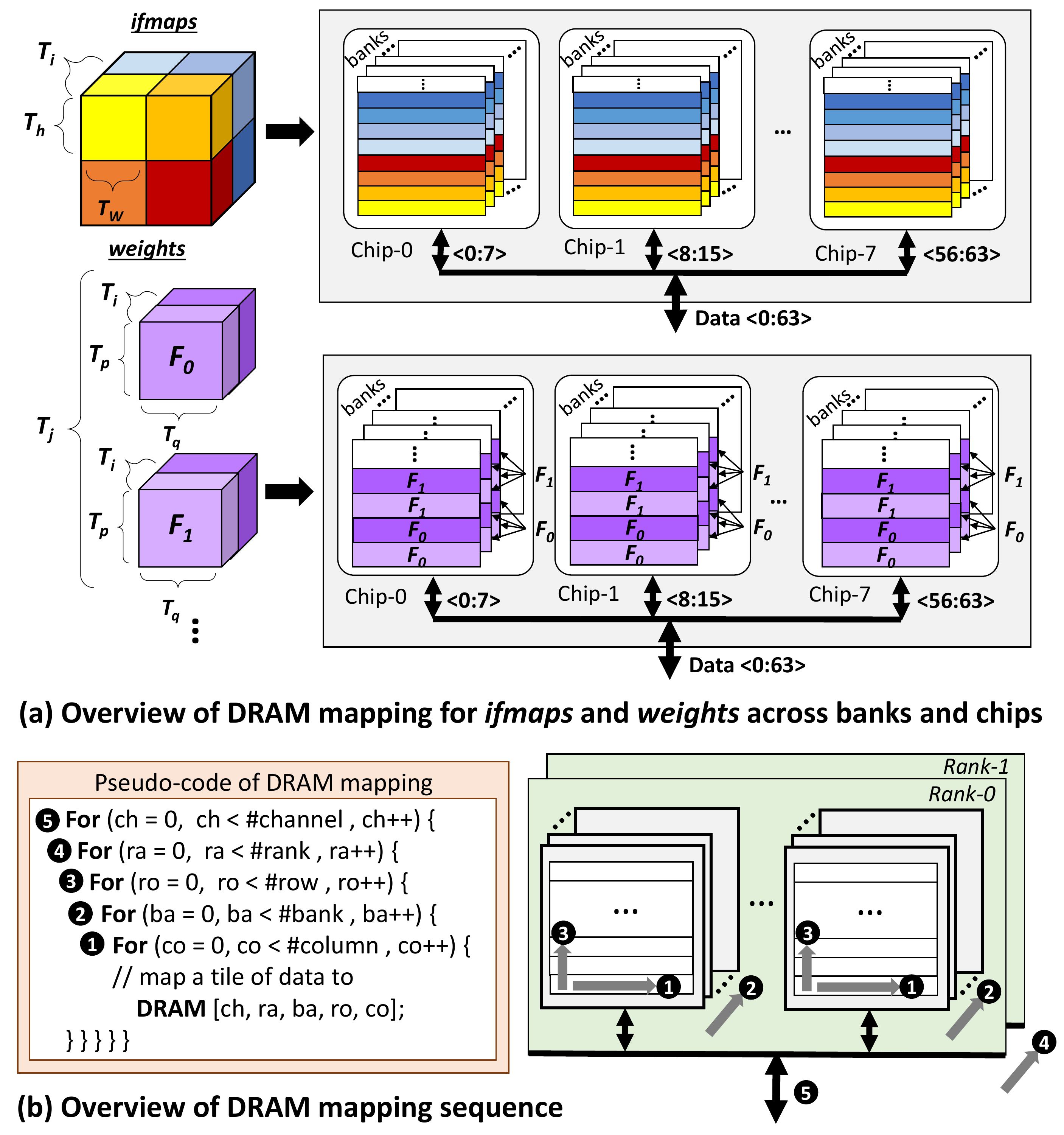}
\vspace{-0.7cm}
\caption{(a) Overview of the proposed DRAM mapping. (b) Proposed DRAM mapping sequence. In the pseudo-code, \#column denotes the number of columns in a row, \#row denotes the number of rows in a bank, \#bank denotes the number of banks in a chip, \#rank denotes the number of ranks in a module, and \#channel denotes the number of channel.}
\label{Fig:MemoryMapping_DRAM}
\vspace{-0.3cm}
\end{figure}

\textbf{Proposed DRAM Data Mapping:} 
Here, multiple data from the same tile which are expected to be accessed subsequently in a short time interval, are placed in the same row of the same bank, thereby increasing the row buffer hits.
To further minimize the row buffer conflicts while increasing the throughput, chip- and bank-level parallelism are exploited.
Exploiting chip-level parallelism means that if multiple data from the same tile are expected to be fetched in parallel, these data are placed across chips, if applicable. 
Similarly, for bank-level parallelism, the data are placed across banks, if applicable.
\add{Our DRAM mapping also exploits the DRAM multi-bank burst feature to further improve the DRAM throughput}. 
This mapping concept is illustrated in Fig. \ref{Fig:MemoryMapping_DRAM}(a). 

We further define the DRAM mapping sequence, based on our observations in Fig. \ref{Fig:DDR3_EnergyLatency}. 
The proposed mapping sequence is shown Fig. \ref{Fig:MemoryMapping_DRAM}(b). 
In step-\circledB{1}, we prioritize to map a data tile to different columns in the same row, to achieve maximum row buffer hits. 
This step can be done across different chips in parallel, if applicable, to exploit chip-level parallelism.
If all columns in the same row are fully filled, then the remaining data are mapped in the different banks in the same chip, to exploit bank-level parallelism (step-\circledB{2}).
Mapping to different banks can also be done across different chips in parallel, if applicable.
For each bank, data are mapped to different columns in the same row, just like step-\circledB{1}.
Here, if all columns in the same row are fully filled, then the remaining data are mapped to a different row (step-\circledB{3}).
These steps \circledB{1}-\circledB{3} are repeated until all data are mapped in a DRAM rank.
If there are remaining data left, they can be mapped in different ranks (step-\circledB{4}) and channels (step-\circledB{5}) respectively if applicable, using the same steps as \circledB{1}-\circledB{3}.

To effectively perform DRAM mapping, the information of layer partitioning and scheduling for different data types is required. 
\add{Note that our DRAM mapping scheme can be employed for all DRAM variants (e.g., DDR3, DDR4, etc.), since all of them have similar internal organization, i.e., they are composed of channels, ranks, chips, banks, rows, and columns, when viewed from a top-down perspective as shown in Fig. \ref{Fig:DRAMorgMultiBank} \cite{Ref:Ghose_DRAMworkload_MACS19}.}

\vspace{-0.2cm}
\subsection{Data Mapping in the SRAM Buffer}
\label{Sec:ROMANet_MappingSPM}
To efficiently transport the data between the DRAM and the compute engine, a data mapping in the SRAM buffer is required. 
Since the DRAM mapping considers tile-wise mapping, hence the SRAM buffer considers the same.
Here, we employs SRAM buffer in the scratch-pad memory fashion, since it is commonly used in many CNN accelerators as the local buffer \cite{Ref:Stoutchinin_Scheduling_arXiv19}.
The idea of the mapping is that, if multiple data from the same tile are expected to be fetched in parallel, these data are placed across banks to exploit bank-level parallelism.
Depending upon the configuration of the SRAM buffer (e.g., capacity and number of banks), we can devise an efficient data mapping. 

\begin{figure}[hbtp]
\vspace{-0.3cm}
\centering
\includegraphics[width=\linewidth]{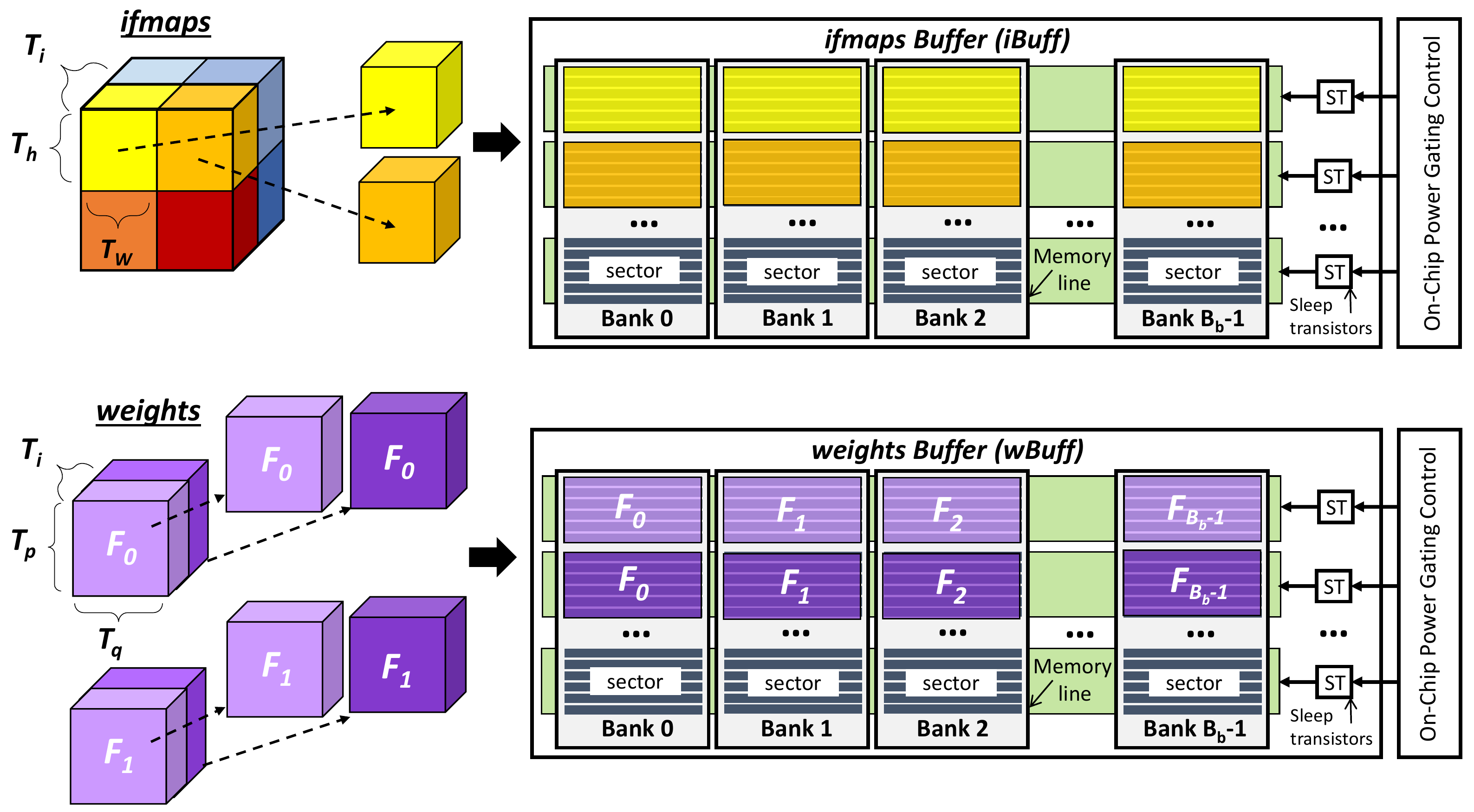}
\vspace{-0.7cm}
\caption{Overview of the concept in SRAM buffer data mapping for \textit{ifmaps} and \textit{weights}, across different banks.}
\label{Fig:MemoryMapping_Buffer}
\vspace{-0.3cm}
\end{figure}

\textbf{Proposed SRAM Buffer Data Mapping:} The concept of the proposed mapping is illustrated in Fig. \ref{Fig:MemoryMapping_Buffer}. 
Here, we prioritize to map a data tile (for each data type) in the same row, across different banks, to achieve maximum bank-level parallelism. 
If the same row across different banks are fully filled, then the remaining data are mapped in a different row, across different banks.
These steps are repeated until all data from the same tile are mapped in the SRAM buffer.
In this manner, each tile may occupy multiple subsequent rows.
For \textit{weights} data type, if systolic array-based CNN accelerator is considered, different filters can be mapped in different banks, thus each bank can supply specific filter(s) to specific column of the systolic array, as illustrated in Fig. \ref{Fig:MemoryMapping_Buffer}.
Furthermore, we can also employ different sectors in the buffer, and each of which has dedicated sleep transistor to power-gate the unused sectors, for obtaining lower power and energy.

\section{Evaluation Methodology}
\label{Sec:EvalMethod}

\begin{figure}[hbtp]
\vspace{-0.3cm}
\centering
\includegraphics[width=\linewidth]{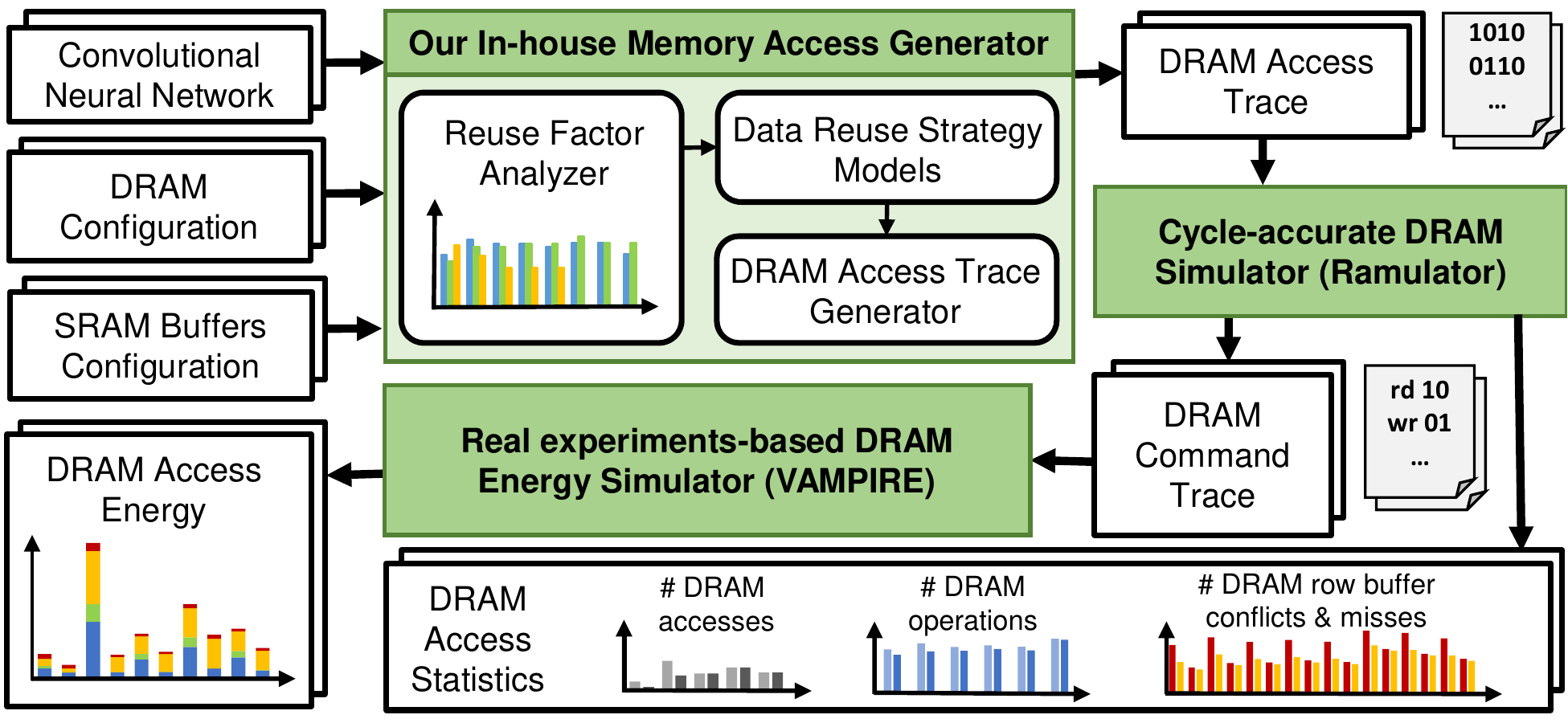}
\vspace{-0.7cm}
\caption{Our experimental setup and tool flow.}
\label{Fig:ToolFlow}
\vspace{-0.3cm}
\end{figure}

We built our experimental setup and tool flow that integrate a memory access generator, a cycle-accurate DRAM simulator, and a real experiments-based DRAM energy simulator (as shown in Fig. \ref{Fig:ToolFlow}). 
\vspace{0.1cm}\\
\textbf{Memory Access Generator:}
Our in-house memory access generator receives the CNN information (e.g., number of layers and data size) and configuration of the DRAM and the SRAM buffers, as the inputs. It produces a DRAM access trace that represents the sequence of DRAM requests, as the output.
Our tool analyzes the characteristics of CNN and extracts its reuse factor information.
The reuse factor is then used for devising data scheduling, depending upon the methodology (e.g., the ROMANet uses a DSE to find the layer partitioning and scheduling).
Based on the data reuse strategy, the tool generates the representative DRAM access trace. %
\vspace{0.2cm}\\
\textbf{Cycle-accurate DRAM Simulator:}
We used a state-of-the-art cycle-accurate DRAM simulator (i.e., Ramulator \cite{Ref:Kim_Ramulator_LCA15}) to simulate the DRAM. 
The input is the DRAM access trace from our memory access generator. 
The outputs are DRAM command trace and statistics, such as the number of accesses, operations, row buffer conflicts and row buffer misses.
\vspace{0.2cm}\\
\textbf{DRAM Energy Simulator:}
To estimate the DRAM access energy for the defined DRAM requests, we used a state-of-the-art real experiments-based DRAM energy simulator (i.e., VAMPIRE \cite{Ref:Ghose_VAMPIRE_POMACS18}). 
The input is the DRAM command trace generated by the DRAM simulator and the output is the DRAM access energy estimation.

\vspace{0.2cm}
In this evaluation, we considered a state-of-the-art Tensor Processing Unit (TPU) \cite{Ref:Jouppi_TPU_ISCA17}-like CNN hardware accelerator, as specified in Fig. \ref{Fig:DNNaccelerator} and Table \ref{Table:Accelerator}. 
We used DRAM with DDR3-1600 2Gb x8 configuration \cite{Ref:Micron} \cite{Ref:Malladi_DRAM_ISCA12}. 
Here, the DRAM \textit{open row} policy was used since it keeps the row open after an access, hence the subsequent accesses to the same row can be done fast and with less energy.
\add{We considered first-come first-serve (FCFS) scheduling policy for handling DRAM requests, which totally disregards the current state of the DRAM row buffer and serves the request that is received first. 
We used different DRAM access modes: (1) \textit{burst mode} with burst length = 8 (BL8) for DDR3 DRAM \cite{Ref:Micron}, which provides multiple words of data per-request, and (2) \textit{non-burst mode}, which provides a single word of data per-request.}
For the CNNs, we used the AlexNet~\cite{Ref:Alex_AlexNet_NIPS12}, the VGG-16~\cite{Ref:Simonyan_VGG16_arXiv14}, and the MobileNet~\cite{Ref:Howard_MobileNet_arXiv17}. 
\begin{table}[hbtp]
\vspace{-0.2cm}
\caption{Configuration of the systolic array-based CNN accelerator.}
\vspace{-0.1cm}
\label{Table:Accelerator}
\centering
\small
\begin{tabular}{|l|l|}
\hline 
\multicolumn{1}{|c|}{\textbf{Module}} & \multicolumn{1}{c|}{\textbf{Description}} \\
\hline
\hline
Systolic Array & Size = 8 $\times$ 8 MACs \\
\hline
\multirow{2}{*}{Buffers} & 64KB \textit{iBuff}, 64KB \textit{wBuff}, 64KB \textit{oBuff}\\
     & $\#$banks = 8 banks-per-buffer\\ 
\hline
\multirow{2}{*}{Memory Controller} &  Policy = open row\\
     & Scheduler = FCFS\\ 
\hline
\multirow{4}{*}{DRAM} & DDR3-1600 2Gb x8\\ 
     & 1 channel, 1 rank-per-channel\\
     & 1 chip-per-rank, 8 banks-per-chip\\
     & Mode: non-burst, burst (burst length = 8)\\
\hline
\end{tabular}
\vspace{-0.2cm}
\end{table}


\begin{figure*}[t]
\centering
\includegraphics[width=\linewidth]{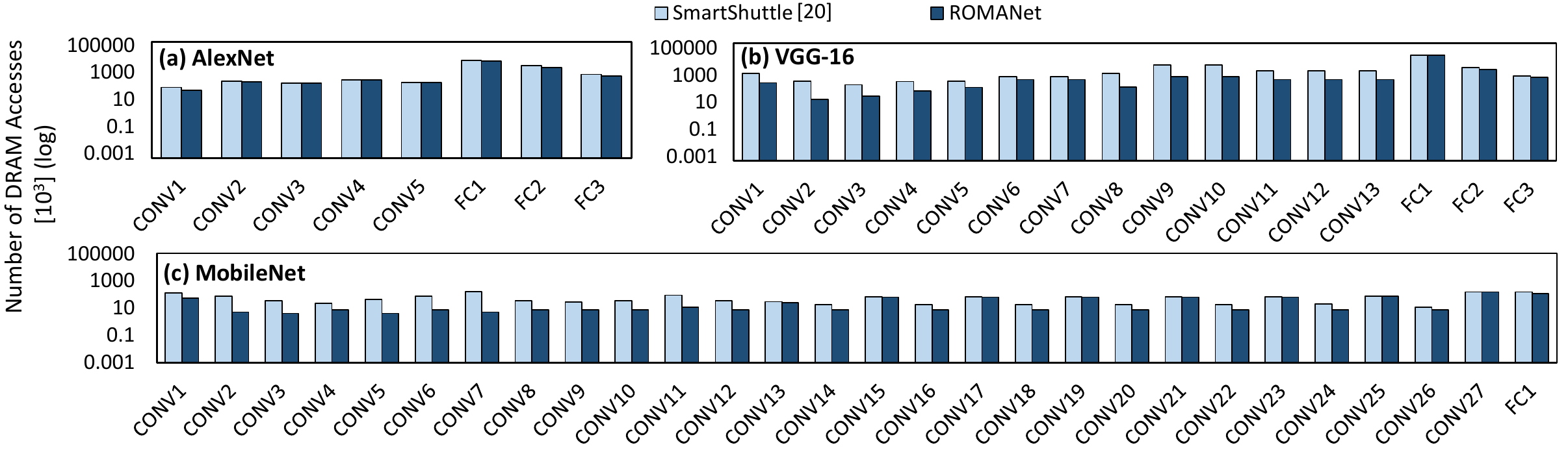}
\vspace{-0.7cm}
\caption{Results for the number of DRAM accesses in (a) AlexNet, (b) VGG-16 and (b) MobileNet.}
\label{Fig:Results_nAccess_DNNs}
\vspace{-0.4cm}
\end{figure*}


To study the improvements offered by the ROMANet over the state-of-the-art (or baseline) of the layer partitioning and scheduling scheme (i.e., SmartShuttle \cite{Ref:Li_SmartShuttle_DATE18}), we implemented the baseline \cite{Ref:Li_SmartShuttle_DATE18} and integrated it in our experimental setup.
The differences between the ROMANet and the baseline are the data reuse strategy and DRAM mapping. 
For data reuse strategy, the baseline uses \textit{ofmaps}- and \textit{weights}-reuse scheduling. 
\add{For DRAM mapping, since the SmartShuttle has no defined DRAM mapping scheme, we used the state-of-the-art mapping scheme from \cite{Ref:Zhang_Caffeine_TCAD19} for the baseline, which maps each data partition (tile) in a continuous address space in a DRAM bank to enable the effective use of the DRAM \textit{burst mode}.}

\section{Results and Discussion}
\label{Sec:Results}

\subsection{Reduction in the Number of DRAM Accesses}
\label{Sec:Results_DRAMaccess}

Evaluation results for the number of DRAM accesses are presented in Fig.~\ref{Fig:Results_nAccess_DNNs}(a) for the AlexNet, Fig.~\ref{Fig:Results_nAccess_DNNs}(b) for the VGG-16, and  Fig.~\ref{Fig:Results_nAccess_DNNs}(c) for the MobileNet. 
ROMANet decreases the number of DRAM accesses by $12\%$ over the baseline for the AlexNet, by $36\%$ for the VGG-16, and by $45\%$ for the MobileNet.
These improvements are achieved because of the effective layer partitioning and scheduling performed by the ROMANet, and can be associated with several aspects.
\begin{enumerate}[leftmargin=*]
    \item The analytical model in the ROMANet considers the overlapping data that do not need to be re-fetched from the DRAM, while the analytical model of the baseline does not consider the overlapping data. 
    \addx{It contributes to 5\% DRAM access reduction in AlexNet, 16\% in VGG-16, and 21\% in MobileNet.}     
    \item The ROMANet defines the layer partitioning by exploring all possible values of tiling factors of all data types in the DSE, while the baseline employs additional rules, e.g., prioritize to maximize tiling factor $T_j$.
    \item The ROMANet considers more possible scheduling schemes than the baseline, as they are devised from reuse factors and reuse priority orders. However, the baseline considers only the \textit{weights}- and \textit{ofmaps}-based scheduling schemes.
    \addx{It contributes to 7\% DRAM access reduction in AlexNet, 20\% in VGG-16, and 24\% in MobileNet.}  
\end{enumerate}
Therefore, the ROMANet considers a wider search space than the baseline, as it investigates the number of DRAM accesses in more detailed analytical model and more number of layer partitioning and scheduling schemes, which in-turn open a higher possibility to find the configuration that leads to less number of DRAM accesses.
These results also show that the reductions of the DRAM accesses happen on layer-wise basis, which is in-line with the defined optimization problem. 

\vspace{-0.6cm}
\add{\subsection{Reduction in the Number of DRAM Requests}}
\label{Sec:Results_DRAMrequest}

\begin{figure}[hbtp]
\vspace{-0.2cm}
\centering
\includegraphics[width=\linewidth]{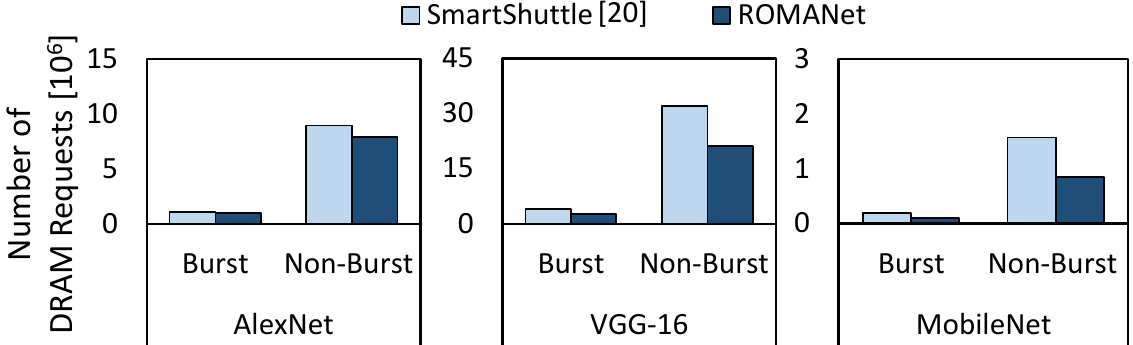}
\vspace{-0.7cm}
\caption{\add{The total number of DRAM requests during an inference with the AlexNet, the VGG-16, and the MobileNet for different DRAM access modes: burst mode and non-burst mode.}}
\label{Fig:Results_nRequest_DNNs}
\vspace{-0.3cm}
\end{figure}

\add{In practice, CNN accelerators typically exploit DRAM burst mode~\cite{Ref:Zhang_Caffeine_TCAD19, Ref:Qiu_GoDeep_FPGA16, Ref:Guan_FPDNN_FCCM17}.
Therefore, we also studied the impact of the burst mode on the number of DRAM requests, compared to non-burst mode. 
Evaluation results are provided in Fig. \ref{Fig:Results_nRequest_DNNs}.} 
\\
\add{The results show that, the burst mode requires less number of DRAM requests as compared to the non-burst mode, for accessing the same number of data from DRAM. 
For instance, employing the burst mode with BL8 can decrease the number of DRAM requests by 8x.
This is important because the burst mode only requires one-time address decoding to fetch multiple data, while the non-burst mode needs to perform address decoding each time a request is issued. 
Therefore, burst mode incurs less access energy than non-burst mode which will be discussed in Section \ref{Sec:Results_DRAMenergy}.}

\vspace{-0.2cm}
\subsection{Reduction in the Number of DRAM Row Buffer Conflicts and Row Buffer Misses}
\label{Sec:Results_DRAMconflicts}

\begin{figure}[hbtp]
\vspace{-0.2cm}
\centering
\includegraphics[width=\linewidth]{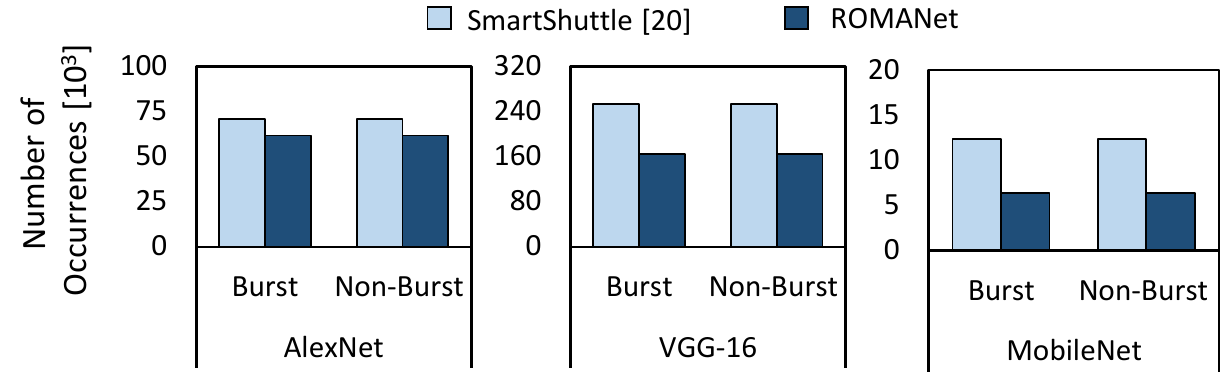}
\vspace{-0.7cm}
\caption{\add{Results for the total number of DRAM row buffer conflicts and misses in (a) AlexNet, (b) VGG-16, and (c) MobileNet.}}
\label{Fig:DRAM_MissesConflicts_DNNs}
\vspace{-0.3cm}
\end{figure}

The evaluation results for the number of DRAM row buffer conflicts and misses are presented in Fig. \ref{Fig:DRAM_MissesConflicts_DNNs} for the AlexNet, the VGG-16, and the MobileNet.
The figure shows that, in general, our ROMANet decreases the number of DRAM row buffer conflicts and misses as compared to the baseline. 
\add{For both burst mode and non-burst mode, ROMANet reduces the row buffer conflicts and misses by about $12\%$ for the AlexNet, $35\%$ for the VGG-16, and $48\%$ for the MobileNet compared to the baseline.
These reductions are achieved because ROMANet devises an effective DRAM mapping policy that minimizes subsequent accesses to different rows in the same DRAM bank, when accessing a single tile partition. 
On the other hand, the addressing in the baseline makes the subsequent accesses to different rows in the same DRAM bank happen more frequent, when accessing a single tile partition. 
Therefore, it has higher possibility of row buffer conflicts than the proposed technique.}

\vspace{-0.2cm}
\subsection{Reduction in the Number of DRAM Operations}
\label{Sec:Results_DRAMops}

\begin{figure}[hbtp]
\vspace{-0.2cm}
\centering
\includegraphics[width=\linewidth]{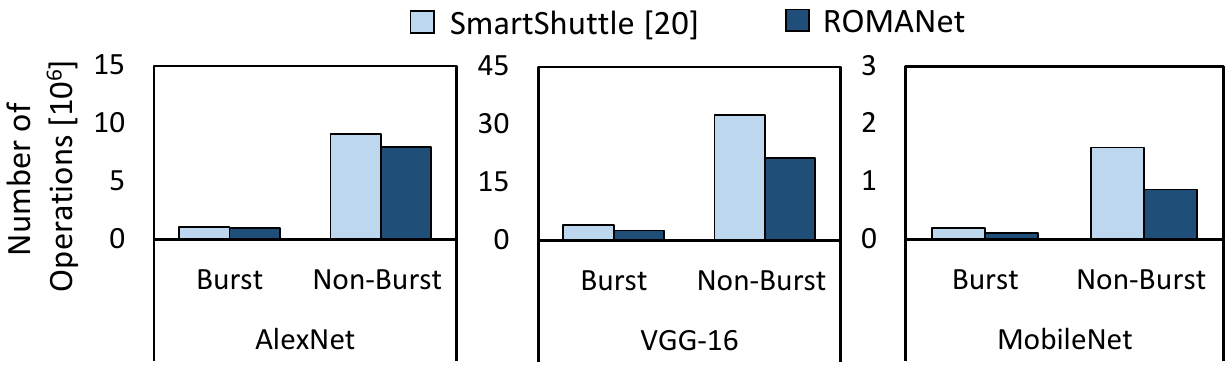}
\vspace{-0.7cm}
\caption{\add{Results for the total number of DRAM operations performed in AlexNet, VGG-16, and MobileNet. 
Each bar includes the number of DRAM \textit{activation}, \textit{precharge}, \textit{read}, and \textit{write}.}}
\label{Fig:Results_Operation_DNNs}
\vspace{-0.3cm}
\end{figure}

The reductions in the number of DRAM requests decrease the total number of DRAM operations, especially for \textit{read} and \textit{write}.
Meanwhile, the reductions in the number of DRAM row buffer conflicts and misses also decrease the total number of DRAM operations, especially for \textit{activation} and \textit{precharge}.
The reason is that, if there is no row buffer conflict or miss, then there is no need for precharging and/or activating a different row in a DRAM bank. 
Moreover, the open row policy keeps the currently activated row open for some time to facilitate subsequent accesses to the same row at a faster rate and with less energy consumption. 
Therefore, the \textit{precharge} and/or \textit{activation} operations are not performed and the number of DRAM operations is reduced.
It is validated by the evaluation results presented in Fig.~\ref{Fig:Results_Operation_DNNs} for the AlexNet, the VGG-16, and the MobileNet.
Each bar of the graph in these figures already includes all DRAM operations: \textit{activation}, \textit{precharge}, \textit{read}, and \textit{write} operations.

\add{Fig. \ref{Fig:Results_Operation_DNNs} shows that employing different DRAM access modes may lead to different number of operations. 
For both in burst mode and non-burst mode, ROMANet reduces the operations by about $12\%$ for the AlexNet, by about $34\%$ for the VGG-16, and by about $45\%$ for the MobileNet.
The burst mode has less number of DRAM operations as compared to the non-burst mode, since it has less number of DRAM requests (as shown in Fig.~\ref{Fig:Results_nRequest_DNNs}).}
The reduction of DRAM operations is important because each operation consumes some energy. 
Therefore, reducing the number of DRAM operations leads to DRAM access energy savings, which will be discussed in Section~\ref{Sec:Results_DRAMenergy}. 

%
\begin{figure*}[t]
\centering
\includegraphics[width=\linewidth]{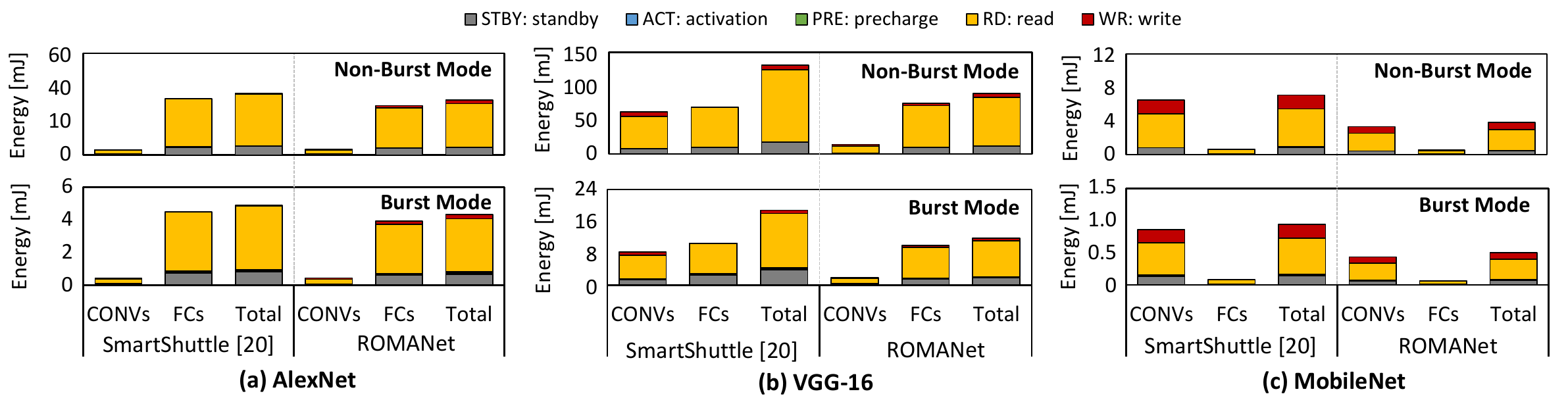}
\vspace{-0.7cm}
\caption{\add{Results for DRAM access energy and the breakdown showing the contribution of different types of DRAM operations in (a) AlexNet, (b) VGG-16, and (c) MobileNet. For AlexNet: CONVs refers to CONV1-CONV5 layers and FCs refers to FC1-FC3 layers; for VGG-16: CONVs refers to CONV1-CONV13 layers and FCs refers to FC1-FC3 layers; and for MobileNet: CONVs refers to CONV1-CONV27 layers and FCs refers to FC1 layer.}}
\label{Fig:Results_DRAM_Energy_DNNs}
\vspace{-0.4cm}
\end{figure*}

\vspace{-0.3cm}
\subsection{DRAM Access Energy Savings}
\label{Sec:Results_DRAMenergy}

The evaluation results for the DRAM access energy are presented in Fig.~\ref{Fig:Results_DRAM_Energy_DNNs} for the AlexNet, the VGG-16, and the MobileNet.
The figure shows that, in general, our ROMANet reduces the DRAM access energy significantly as compared to the baseline, i.e., by about $12\%$ for the AlexNet, $36\%$ for the VGG-16, and $46\%$ for the MobileNet.
The burst mode has less DRAM access energy as compared to the non-burst mode (about 8x energy saving), since it has less number of DRAM operations. 
The main sources of these savings are: 1) the reduction in the number of accesses, and 2) the reduction in the number of row buffer conflicts and misses.

The energy savings from the reduction of accesses can be observed in the figure from the reduction in the energy consumption incurred by the \textit{read} and \textit{write} operations. 
Here, \textit{read} and \textit{write} operations already include operations for all data types (i.e., \textit{ifmaps}, \textit{ofmaps}, and \textit{weights}).
The savings are achieved because of the effective layer partitioning and scheduling found using the design space exploration. 
Meanwhile, the energy savings from the reduction of row buffer conflicts and misses, are reflected by the reduction in the energy consumption incurred by the \textit{activation} and \textit{precharge} operations.
These savings are mainly because of the effective DRAM mapping that exploits row buffer locality, to decrease the row buffer conflicts and misses.  
Furthermore, since the proposed mapping in ROMANet also optimizes the DRAM access latency, the standby energy is also reduced. 

We also observed that, for the AlexNet and the VGG-16, the FC layers consume more access energy compared to the CONV layers, as shown in Fig. \ref{Fig:Results_DRAM_Energy_DNNs}(a)-(b).
\add{This is due to the fact that, in these networks, the FC layers have a large number of \textit{weights} compared to the CONV layers.  
Therefore, to access these weights, a proportional number of DRAM accesses are required, which lead to the high energy consumption shown in the figure. 
Meanwhile in the MobileNet, the FC layer does not dominate the DRAM access energy since the network only has a single FC layer whose number of weights and feature maps are relatively small, i.e., 1024x1000 \textit{weights}, 1024 \textit{ifmaps}, and 1000 \textit{ofmaps}.}

\vspace{-0.7cm}
\add{\subsection{DRAM Access Energy Savings in Sparse CNNs}}
\label{Sec:Results_DRAMenergyPruned}

%

\begin{figure}[hbtp]
\centering
\includegraphics[width=\linewidth]{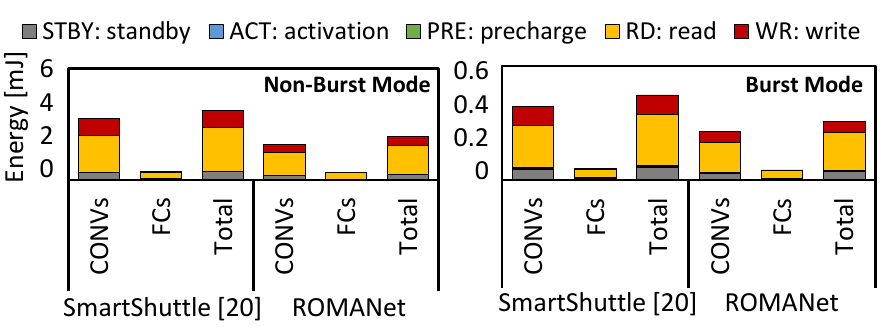}
\vspace{-0.7cm}
\caption{\add{Results for the DRAM access energy and the breakdown showing the contribution of different DRAM operations in the sparse MobileNet.
The CONVs refers to CONV1-CONV27 layers and FCs refers to FC1 layer.
}}
\label{Fig:Results_Energy_SparseMobileNet}
\end{figure}

\add{To improve the energy efficiency of DNN-based systems, DNNs are usually first passed through a compression framework for achieving a compact model that can be deployed in resource constraint mobile devices~\cite{Ref:He_AMC_ECCV18}\cite{han2015deep}. Based on the study of different compression techniques, it has been observed that structured pruning techniques are more commonly employed due to their easy deployability using off-the-shelf libraries~\cite{Ref:He_AMC_ECCV18}\cite{yu2017scalpel}. Therefore, to show the applicability of ROMANet for structurally pruned DNNs, we evaluated ROMANet for a sparse MobileNet pruned using one of the state-of-the-art structured pruning technique, i.e., AutoML for Model Compression (AMC)~\cite{Ref:He_AMC_ECCV18}. 
Fig.~\ref{Fig:Results_Energy_SparseMobileNet} shows the DRAM access energy per inference for the sparse MobileNet. 
The figure shows that our ROMANet reduces the DRAM access energy by about $30\%$ in the burst mode and by about $38\%$ in the non-burst mode, when compared to their respective baselines. 
The main contributors to these savings are also the reductions in the number of DRAM accesses and the number of DRAM row buffer conflicts and misses.} 
\add{These evaluation results prove that our ROMANet can provide DRAM access energy savings even for sparse CNNs when compared to the state-of-the-art technique.}

\vspace{-0.2cm}
\add{\subsection{Data Throughput Improvement for DRAM Accesses}}
\label{Sec:Results_Throughput}

\begin{figure}[hbtp]
\vspace{-0.2cm}
\centering
\includegraphics[width=\linewidth]{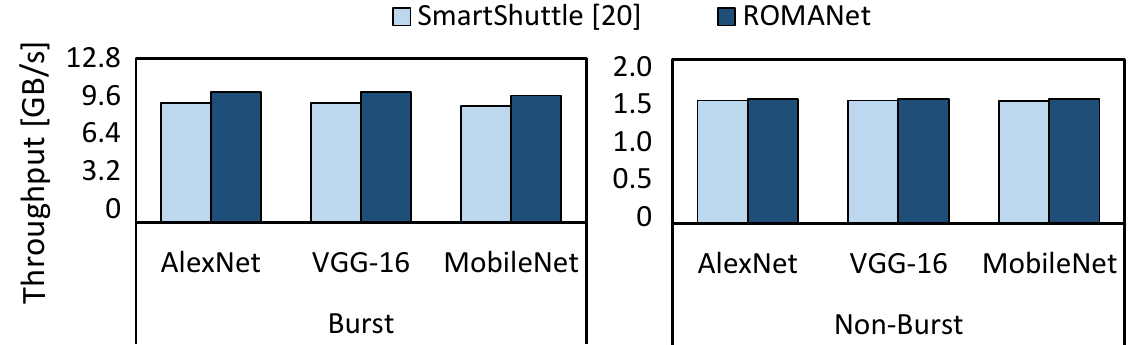}
\vspace{-0.7cm}
\caption{\add{Results for the data throughput of DRAM accesses for AlexNet, VGG-16, and MobileNet using DDR3-1600 DRAM.}}
\label{Fig:Results_Throughput}
\end{figure}

\add{Besides optimizing DRAM access energy, our ROMANet also improves the DRAM throughput in CNN accelerators.
This can be observed from the experimental results presented in Fig.~\ref{Fig:Results_Throughput}. 
The figure shows that ROMANet improves the data throughput by about $10\%$ in burst mode and $1.5\%$ in non-burst mode, compared to the respective baselines.
The main source of the improvement is that the proposed DRAM mapping uses bank-level parallelism to exploit DRAM multi-bank burst feature, thereby decreasing the possibility of facing row buffer conflicts.  
Fig.~\ref{Fig:Results_Throughput} also shows that throughput improvement in the burst mode is typically higher than the non-burst mode. 
The reason is that the non-burst mode requires higher number of DRAM operations than the burst mode for accessing the same amount of data, thereby consuming higher DRAM access latency and decreasing the benefit of exploiting DRAM multi-bank burst feature. 
Therefore, DRAM burst mode is preferred in practical implementation of CNN accelerators.}

\add{DRAM throughput improvement achieved by ROMANet is important because it enables fast data transfer between the off-chip DRAM and the on-chip SRAM buffers of a DNN accelerator. 
This is beneficial for most of the existing accelerators, such as \cite{Ref:Jouppi_TPU_ISCA17}\cite{Ref:Lu_FlexFlow_HPCA17}, because the obtained DRAM throughput can fulfill the bandwidth requirement to fully utilize the on-chip compute engines, thereby maximizing their processing potential.}  
 
\add{In general, all evaluation results (presented in Section~\ref{Sec:Results_DRAMaccess} - Section~\ref{Sec:Results_Throughput}) show that the ROMANet methodology can provide significant DRAM access energy savings while offering a high DRAM throughput for CNN accelerators as compared to the state-of-the-art work.} 

\section{Conclusion}
\label{Sec:Conclusion}
We propose a novel ROMANet methodology to reduce the DRAM access energy \add{and to improve the DRAM throughput} for CNN accelerators. 
It leverages the knowledge of the reuse factor from a given network configuration to perform a design space exploration that finds the effective layer partitioning and scheduling that offer minimum DRAM accesses.
\add{It also exploits DRAM row buffer locality and DRAM multi-bank burst feature by employing effective DRAM data mapping.}
The experimental results prove that the ROMANet can reduce the number of the DRAM accesses, the row buffer conflicts and misses, as well as the operations that lead to significant DRAM access energy savings \add{while improving the data throughput}, as compared to the state-of-the-art.
These results also demonstrate that a synergistic design of layer partitioning and scheduling with data mapping in memory, is important to achieve substantial energy-efficiency \add{and throughput} improvements for DRAM accesses in the CNN accelerators. 
Furthermore, our novel concepts would enable further studies on energy-efficient CNN accelerators. 







\ifCLASSOPTIONcaptionsoff
  \newpage
\fi



%

\bibliographystyle{IEEEtran}
\bibliography{reference}

\begin{thebibliography}{10}
\providecommand{\url}[1]{#1}
\csname url@samestyle\endcsname
\providecommand{\newblock}{\relax}
\providecommand{\bibinfo}[2]{#2}
\providecommand{\BIBentrySTDinterwordspacing}{\spaceskip=0pt\relax}
\providecommand{\BIBentryALTinterwordstretchfactor}{4}
\providecommand{\BIBentryALTinterwordspacing}{\spaceskip=\fontdimen2\font plus
\BIBentryALTinterwordstretchfactor\fontdimen3\font minus
  \fontdimen4\font\relax}
\providecommand{\BIBforeignlanguage}[2]{{%
\expandafter\ifx\csname l@#1\endcsname\relax
\typeout{** WARNING: IEEEtran.bst: No hyphenation pattern has been}%
\typeout{** loaded for the language `#1'. Using the pattern for}%
\typeout{** the default language instead.}%
\else
\language=\csname l@#1\endcsname
\fi
#2}}
\providecommand{\BIBdecl}{\relax}
\BIBdecl

\bibitem{Ref:LeCun_DeepLearning_Nature15}
Y.~LeCun \emph{et~al.}, ``Deep learning,'' \emph{nature}, vol. 521, no. 7553,
  2015.

\bibitem{Ref:Chen_DianNao_ASPLOS14}
T.~Chen \emph{et~al.}, ``Diannao: A small-footprint high-throughput accelerator
  for ubiquitous machine-learning,'' in \emph{Proc. of the 19th Int. Conf. on
  Architectural Support for Programming Languages and Operating Systems
  (ASPLOS)}, 2014, pp. 269--284.

\bibitem{Ref:Zhang_CNNfpga_FPGA15}
C.~Zhang \emph{et~al.}, ``Optimizing fpga-based accelerator design for deep
  convolutional neural networks,'' in \emph{Proc. of the ACM/SIGDA Int. Symp.
  on Field-Programmable Gate Arrays (FPGA)}, 2015, pp. 161--170.

\bibitem{Ref:Han_EIE_ISCA16}
S.~{Han} \emph{et~al.}, ``Eie: Efficient inference engine on compressed deep
  neural network,'' in \emph{Proc. of the ACM/IEEE 43rd Annual Int. Symp. on
  Computer Architecture (ISCA)}, June 2016, pp. 243--254.

\bibitem{Ref:Chen_Eyeriss_JSSC16}
Y.~{Chen} \emph{et~al.}, ``Eyeriss: An energy-efficient reconfigurable
  accelerator for deep convolutional neural networks,'' \emph{IEEE Journal of
  Solid-State Circuits (JSSC)}, vol.~52, no.~1, pp. 127--138, Jan. 2017.

\bibitem{Ref:Zhang_CambriconX_MICRO15}
S.~{Zhang} \emph{et~al.}, ``Cambricon-x: An accelerator for sparse neural
  networks,'' in \emph{Proc. of the IEEE/ACM 49th Annual Int. Symp. on
  Microarchitecture (MICRO)}, Oct. 2016, pp. 1--12.

\bibitem{Ref:Albericio_Cnvlutin_ISCA16}
J.~{Albericio} \emph{et~al.}, ``Cnvlutin: Ineffectual-neuron-free deep neural
  network computing,'' in \emph{Proc. of the ACM/IEEE 43rd Annual Int. Symp. on
  Computer Architecture (ISCA)}, June 2016, pp. 1--13.

\bibitem{Ref:Luo_DaDianNao_TC17}
T.~{Luo} \emph{et~al.}, ``Dadiannao: A neural network supercomputer,''
  \emph{IEEE Transactions on Computers (TC)}, vol.~66, no.~1, pp. 73--88, Jan.
  2017.

\bibitem{Ref:Jouppi_TPU_ISCA17}
N.~P. {Jouppi} \emph{et~al.}, ``In-datacenter performance analysis of a tensor
  processing unit,'' in \emph{Proc. of the ACM/IEEE 44th Annual Int. Symp. on
  Computer Architecture (ISCA)}, June 2017, pp. 1--12.

\bibitem{Ref:Parashar_SCNN_ISCA17}
A.~{Parashar} \emph{et~al.}, ``Scnn: An accelerator for compressed-sparse
  convolutional neural networks,'' in \emph{Proc. of the ACM/IEEE 44th Annual
  Int. Symp. on Computer Architecture (ISCA)}, June 2017, pp. 27--40.

\bibitem{Ref:Lu_FlexFlow_HPCA17}
W.~{Lu} \emph{et~al.}, ``Flexflow: A flexible dataflow accelerator architecture
  for convolutional neural networks,'' in \emph{Proc. of the IEEE Int. Symp. on
  High Performance Computer Architecture (HPCA)}, Feb. 2017, pp. 553--564.

\bibitem{Ref:Kwon_MAERI_ASPLOS18}
H.~Kwon \emph{et~al.}, ``Maeri: Enabling flexible dataflow mapping over dnn
  accelerators via reconfigurable interconnects,'' in \emph{Proc. of the 23th
  Int. Conf. on Architectural Support for Programming Languages and Operating
  Systems (ASPLOS)}, 2018, pp. 461--475.

\bibitem{Ref:Hanif_MPNA_arXiv18}
M.~A. Hanif \emph{et~al.}, ``{MPNA:} {A} massively-parallel neural array
  accelerator with dataflow optimization for convolutional neural networks,''
  \emph{CoRR}, vol. abs/1810.12910, 2018.

\bibitem{Ref:Sharma_BitFusion_ISCA18}
H.~{Sharma} \emph{et~al.}, ``Bit fusion: Bit-level dynamically composable
  architecture for accelerating deep neural network,'' in \emph{Proc. of the
  ACM/IEEE 45th Annual Int. Symp. on Computer Architecture (ISCA)}, June 2018,
  pp. 764--775.

\bibitem{Ref:Kim_Ramulator_LCA15}
Y.~{Kim} \emph{et~al.}, ``Ramulator: A fast and extensible dram simulator,''
  \emph{IEEE Computer Architecture Letters (LCA)}, vol.~15, no.~1, pp. 45--49,
  Jan. 2016.

\bibitem{Ref:Ghose_VAMPIRE_POMACS18}
S.~Ghose \emph{et~al.}, ``What your dram power models are not telling you:
  Lessons from a detailed experimental study,'' \emph{Proc. of ACM Measurement
  and Analysis of Computing Systems}, vol.~2, no.~3, pp. 38:1--38:41, Dec.
  2018.

\bibitem{Ref:Sze_DNNsurvey_IEEE17}
V.~{Sze} \emph{et~al.}, ``Efficient processing of deep neural networks: A
  tutorial and survey,'' \emph{Proceedings of the IEEE}, vol. 105, no.~12, pp.
  2295--2329, Dec 2017.

\bibitem{Ref:Horowitz_ComputeEnergy_ISSCC14}
M.~{Horowitz}, ``1.1 computing's energy problem (and what we can do about
  it),'' in \emph{Proc. of the IEEE Int. Solid-State Circuits Conf. Digest of
  Technical Papers (ISSCC)}, Feb. 2014, pp. 10--14.

\bibitem{Ref:Hedge_UCNN_ISCA18}
K.~Hegde \emph{et~al.}, ``Ucnn: Exploiting computational reuse in deep neural
  networks via weight repetition,'' in \emph{Proc. of the 45th Annual Int.
  Symp. on Computer Architecture (ISCA)}, 2018, pp. 674--687.

\bibitem{Ref:Li_SmartShuttle_DATE18}
J.~{Li} \emph{et~al.}, ``Smartshuttle: Optimizing off-chip memory accesses for
  deep learning accelerators,'' in \emph{Prof. of the Design, Automation Test
  in Europe Conference Exhibition (DATE)}, March 2018, pp. 343--348.

\bibitem{Ref:Zhang_Caffeine_TCAD19}
C.~{Zhang} \emph{et~al.}, ``Caffeine: Toward uniformed representation and
  acceleration for deep convolutional neural networks,'' \emph{IEEE Trans. on
  Computer-Aided Design of Integrated Circuits and Systems (TCAD)}, vol.~38,
  no.~11, pp. 2072--2085, Nov. 2019.

\bibitem{Ref:Qiu_GoDeep_FPGA16}
J.~{Qiu} \emph{et~al.}, ``Going deeper with embedded fpga platform for
  convolutional neural network,'' in \emph{Proc. of the ACM/SIGDA Int. Symp. on
  Field-Programmable Gate Arrays (FPGA)}, 2016, pp. 26--35.

\bibitem{Ref:Guan_FPDNN_FCCM17}
Y.~{Guan} \emph{et~al.}, ``Fp-dnn: An automated framework for mapping deep
  neural networks onto fpgas with rtl-hls hybrid templates,'' in \emph{Proc. of
  the IEEE 25th Annual Int. Symp. on Field-Programmable Custom Computing
  Machines (FCCM)}, April 2017, pp. 152--159.

\bibitem{Ref:Wolf_Loop_TPDS91}
M.~E. {Wolf} and M.~S. {Lam}, ``A loop transformation theory and an algorithm
  to maximize parallelism,'' \emph{IEEE Trans. on Parallel and Distributed
  Systems (TPDS)}, vol.~2, no.~4, pp. 452--471, Oct. 1991.

\bibitem{Ref:Akin_DesignSpace_ASAP14}
B.~{Akın} \emph{et~al.}, ``Understanding the design space of dram-optimized
  hardware fft accelerators,'' in \emph{Proc. of the IEEE 25th Int. Conf. on
  Application-Specific Systems, Architectures and Processors (ASAP)}, June
  2014, pp. 248--255.

\bibitem{Ref:Ghose_DRAMworkload_MACS19}
S.~Ghose \emph{et~al.}, ``Demystifying complex workload-dram interactions: An
  experimental study,'' \emph{Proc. of ACM Measurement and Analysis of
  Computing Systems}, vol.~3, no.~3, Dec. 2019.

\bibitem{Ref:Stoutchinin_Scheduling_arXiv19}
A.~Stoutchinin \emph{et~al.}, ``Optimally scheduling {CNN} convolutions for
  efficient memory access,'' \emph{CoRR}, vol. abs/1902.01492, 2019.

\bibitem{Ref:Chen_EyerissV2_arXiv18}
Y.~{Chen} \emph{et~al.}, ``Eyeriss v2: A flexible accelerator for emerging deep
  neural networks on mobile devices,'' \emph{IEEE Journal on Emerging and
  Selected Topics in Circuits and Systems (JETCAS)}, vol.~9, no.~2, pp.
  292--308, June 2019.

\bibitem{Ref:Micron}
Micron, ``Micron 2gb: x4, x8, x16 ddr3 sdram. datasheet mt41j128m16ha-12,''
  2010.

\bibitem{Ref:Malladi_DRAM_ISCA12}
K.~T. {Malladi} \emph{et~al.}, ``Towards energy-proportional datacenter memory
  with mobile dram,'' in \emph{Proc. of the 39th Annual Int. Symp. on Computer
  Architecture (ISCA)}, June 2012, pp. 37--48.

\bibitem{Ref:Alex_AlexNet_NIPS12}
A.~Krizhevsky \emph{et~al.}, ``Imagenet classification with deep convolutional
  neural networks,'' in \emph{Proc. of the Advances in Neural Information
  Processing Systems (NIPS)}, 2012, pp. 1097--1105.

\bibitem{Ref:Simonyan_VGG16_arXiv14}
K.~Simonyan and A.~Zisserman, ``Very deep convolutional networks for
  large-scale image recognition,'' \emph{arXiv preprint arXiv:1409.1556}, 2014.

\bibitem{Ref:Howard_MobileNet_arXiv17}
A.~G. Howard \emph{et~al.}, ``Mobilenets: Efficient convolutional neural
  networks for mobile vision applications,'' \emph{CoRR}, vol. abs/1704.04861,
  2017.

\bibitem{Ref:He_AMC_ECCV18}
Y.~{He} \emph{et~al.}, ``Amc: Automl for model compression and acceleration on
  mobile devices,'' in \emph{Proc. of the European Conference on Computer
  Vision (ECCV)}, 2018, pp. 784--800.

\bibitem{han2015deep}
S.~Han \emph{et~al.}, ``Deep compression: Compressing deep neural networks with
  pruning, trained quantization and huffman coding,'' \emph{arXiv preprint
  arXiv:1510.00149}, 2015.

\bibitem{yu2017scalpel}
J.~{Yu} \emph{et~al.}, ``Scalpel: Customizing dnn pruning to the underlying
  hardware parallelism,'' \emph{ACM SIGARCH Computer Architecture News},
  vol.~45, no.~2, pp. 548--560, 2017.

\end{thebibliography}


%

\vspace{-0.7cm}
\begin{IEEEbiography}[{\includegraphics[width=1in,height=1.25in,clip,keepaspectratio]{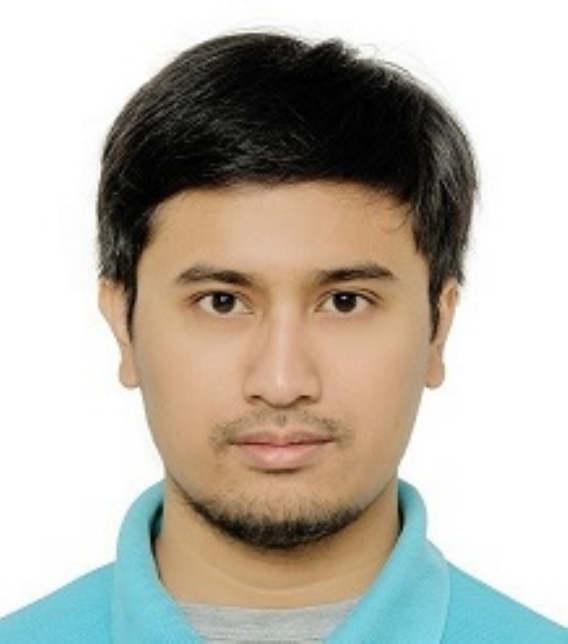}}]
{Rachmad Vidya Wicaksana Putra}
is a research assistant and Ph.D. student at Computer Architecture and Robust Energy-Efficient Technologies (CARE-Tech), Institute of Computer Engineering, Technische Universität Wien (TU Wien), Austria. He received B.Sc. on Electrical Engineering in 2012 and M.Sc. on Electronics in 2015 with distinction, both from Bandung Institute of Technology (ITB), Indonesia. 
He was also a teaching assistant at Electrical Engineering Department ITB in 2012-2017
and as a research assistant at Microelectronics Center ITB in 2014-2017. 
He is a recipient of the Indonesian Endowment Fund for Education (IEFE/LPDP) Scholarship from Ministry of Finance, Indonesia. 
His research interests mainly include computer architecture, VLSI design, system-on-chip, brain-inspired and neuromorphic computing, and electronic design automation.
\end{IEEEbiography}

\vspace{-0.7cm}
\begin{IEEEbiography}[{\includegraphics[width=1in,height=1.25in,clip,keepaspectratio]{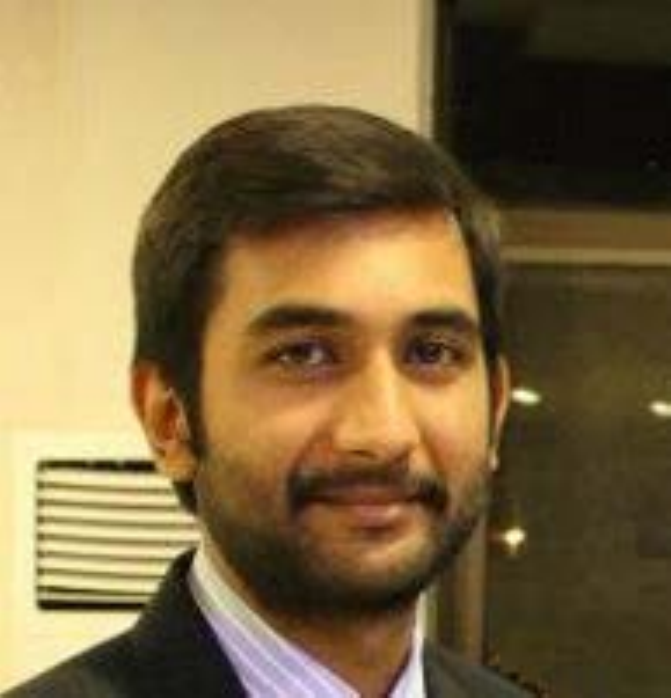}}]
{Muhammad Abdullah Hanif}
is a University Assistant in Vienna University of Technology (TU Wien), Austria. In the past, he has also worked as a Research Associate in Vision Processing (VISpro) lab, Information Technology University (ITU), Pakistan, and as a Lab Engineer in Ghulam Ishaq Khan Institute of Engineering Sciences and Technology (GIKI), Pakistan. He currently holds a Master of Science degree in Electrical Engineering with specialization in Digital Systems and Signal Processing (DSSP) from School of Electrical Engineering and Computer Science (SEECS), National University of Sciences and Technology (NUST), Islamabad, Pakistan, and a Bachelor of Science degree in Electronic Engineering from GIKI, Pakistan. He is also a recipient of President’s Gold Medal for his outstanding academic performance during his MS degree.
\end{IEEEbiography}

\vspace{-0.7cm}
\begin{IEEEbiography}[{\includegraphics[width=1in,height=1.25in,clip,keepaspectratio]{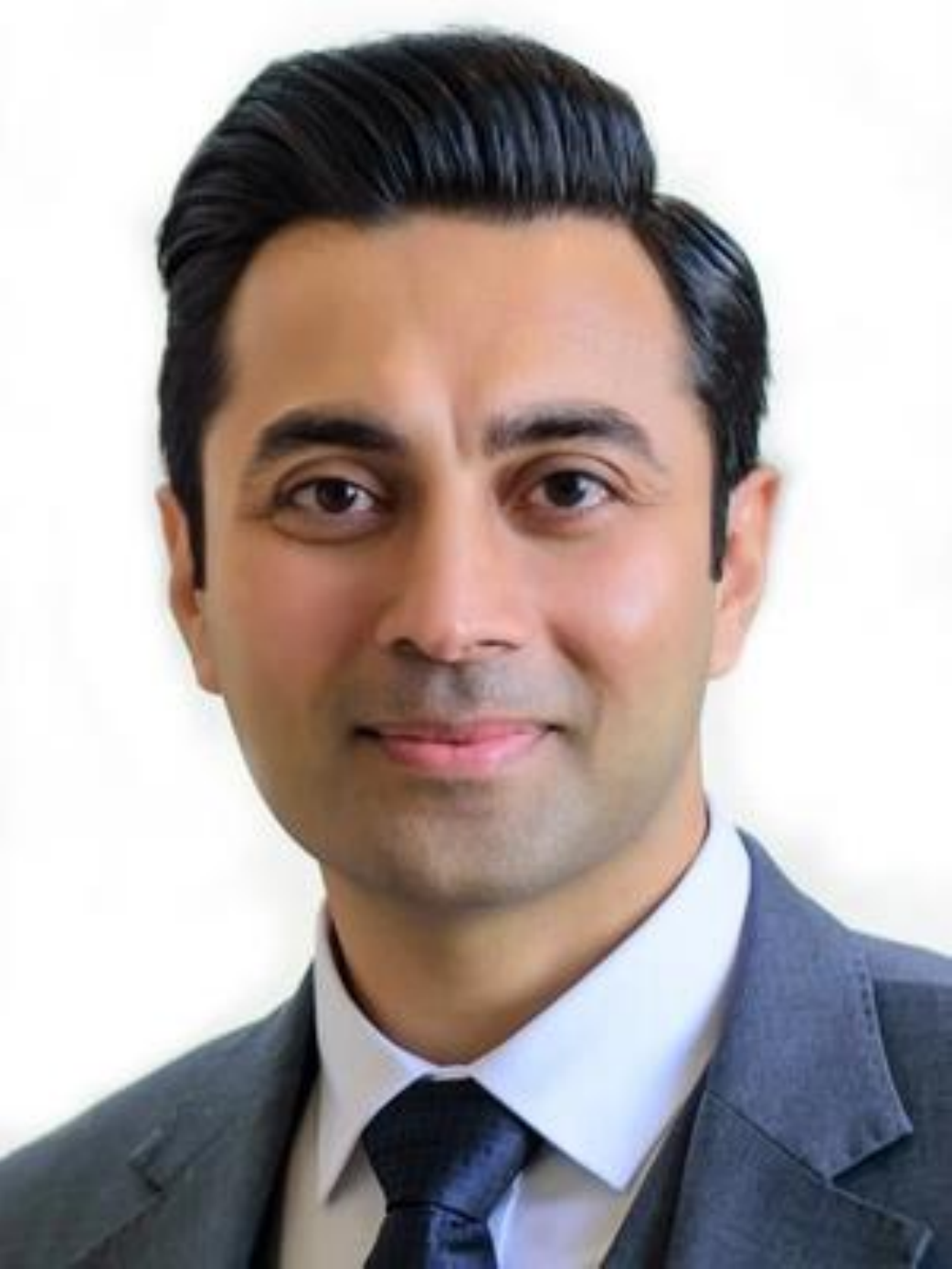}}]
{Muhammad Shafique}

(M’11 - SM’16) is a full professor of Computer Architecture and Robust Energy-Efficient Technologies (CARE-Tech.) at the Institute of Computer Engineering, TU Wien, Austria since Nov. 2016. He received his Ph.D. in Computer Science from Karlsruhe Institute of Technology (KIT), Germany, in Jan.2011. Before, he was with Streaming Networks Pvt. Ltd. where he was involved in research and development of video coding systems for several years. His research interests are in computer architecture, power-/energy-efficient systems, robust computing, hardware security, Brain-Inspired computing trends like Neuromorphic and Approximate Computing, hardware and system-level design for Machine Learning and AI, emerging technologies \& nanosystems, FPGAs, MPSoCs, and embedded systems. His research has a special focus on cross-layer modeling, design, and optimization of computing and memory systems, as well as their deployment in use cases from Internet-of-Things (IoT), Cyber-Physical Systems (CPS), and ICT for Development (ICT4D) domains.

Dr. Shafique has given several Keynote, Invited Talks, and Tutorials. He has also organized many special sessions at premier venues and served as the Guest Editor for IEEE Design and Test Magazine and IEEE Transactions on Sustainable Computing. He has served on the TPC of numerous prestigious IEEE/ACM conferences. Dr. Shafique received the 2015 ACM/SIGDA Outstanding New Faculty Award, six gold medals in his educational career, and several best paper awards and nominations at prestigious conferences like CODES+ISSS, DATE, DAC and ICCAD, Best Master Thesis Award, DAC'14 Designer Track Best Poster Award, IEEE Transactions of Computer "Feature Paper of the Month" Awards, and Best Lecturer Award. Dr. Shafique holds one US patent and has (co-)authored 6 Books, 10+ Book Chapters, and over 200 papers in premier journals and conferences. He is a senior member of the IEEE and IEEE Signal Processing Society (SPS), and a member of the ACM, SIGARCH, SIGDA, SIGBED, and HIPEAC.

\end{IEEEbiography}








\end{document}